\documentclass{tMPH2e}
\usepackage{graphics}
\usepackage{dcolumn}
\usepackage[version=3]{mhchem}
\usepackage[english]{babel}
\usepackage{graphicx}
\usepackage{amsmath}
\usepackage{tabularx}
\usepackage{bigstrut}
\usepackage{longtable}
\usepackage{rotating}
\usepackage{comment}
\usepackage{float}
\usepackage{verbatim}
\usepackage{color, soul}
\usepackage{threeparttable}%The solution to the footnote problem in the tabular environment. Can be used inside a table environment. Make sure the caption is placed inside the table environment and not the threeparttable environment, otherwise the caption is made the same width as the table and the \centering command does not work.
\usepackage{threeparttablex}%The solution to the footnote problem in the longtable environment. This seemed to work well, although the environment is slightly different to the normal threeparttable environment! Also had some problems with the table when it is not split over a page! It seems to want to put a header and a footer on the previous page?! The only solution I could find was to place the table in the text where it is split for sure!
\usepackage[np]{numprint}
\npdecimalsign{\ensuremath{.}}
\npthousandsep{\,}
\npproductsign{\times}
\npfourdigitnosep
\usepackage{lscape}

\newcommand*{\wn}{cm$^{-1}$}
\newcommand*{\AX}{A$^1\Pi -\textrm{X}^1\Sigma^+$}
\newcommand*{\atX}{a$^3\Pi -\textrm{X}^1\Sigma^+$}
\newcommand*{\dX}{d$^3\Delta-\textrm{X}^1\Sigma^+$}
\newcommand*{\eX}{e$^3\Sigma^--\textrm{X}^1\Sigma^+$}
\newcommand*{\apX}{a$'^3\Sigma^+-\textrm{X}^1\Sigma^+$}

\newcommand*{\OSX}{1$^1\Sigma^+-\textrm{X}^1\Sigma^+$}
\newcommand*{\OPX}{1$^1\Pi-\textrm{X}^1\Sigma^+$}
\newcommand*{\BA}{${\rm B}^1\Sigma^+-{\rm A}^1\Pi$}
\newcommand*{\Be}{B$^1\Sigma^+-{\rm e}^3\Sigma^-$}
\newcommand*{\Bd}{B$^1\Sigma^+-{\rm d}^3\!\Delta$}
\newcommand*{\Bap}{B$^1\Sigma^+-{\rm a}'^3\Sigma^+$}
\newcommand*{\BX}{B$^1\Sigma^+-\textrm{X}^1\Sigma^+$}

\newcommand*{\CA}{C$^1\Sigma^+-{\rm A}^1\Pi$}

\newcommand*{\ds}{d$^3\!\Delta$}
\newcommand*{\es}{e$\,^3\Sigma^-$}
\newcommand*{\aps}{a$'^3\Sigma^+$}
\newcommand*{\Ds}{D$\,^1\!\Delta$}
\newcommand*{\Is}{I$\,^1\Sigma^-$}
\newcommand*{\As}{A$^1\Pi$}
\newcommand*{\ats}{a$^3\Pi$}
\newcommand*{\Bs}{B$^1\Sigma^+$}
\newcommand*{\Xs}{X$^1\Sigma^+$}
\newcommand*{\CO}{$^{13}$C$^{18}$O}
\newcommand*{\0}{$v=0$}

\begin{document}

\title{Precision spectroscopy and comprehensive analysis of perturbations in the A$\boldsymbol{^1\Pi(v=0)}$ state of $\boldsymbol{^{13}}$C$\boldsymbol{^{18}}$O}

\author{
R.~Hakalla,$^{a}$
T.~M.~Trivikram,$^{b}$
A.~N.~Heays,$^{c,d}$
E.~J.~Salumbides,$^{b}$
N.~de~Oliveira,$^{e}$
R.~W.~Field,$^{f}$ and
W.~Ubachs$^{b}$
\\\vspace{6pt}%
$^{a}${\em{Materials Spectroscopy Laboratory, Faculty of Mathematics and Natural Science, University of Rzesz\'{o}w, Pigonia 1 Street, 35-959 Rzesz\'{o}w, Poland}}\\%
$^{b}${\em{Department of Physics and Astronomy, LaserLaB, Vrije Universiteit, De Boelelaan 1081, 1081 HV Amsterdam, The Netherlands}}\\%
$^{c}${\em{LERMA, Observatoire de Paris, PSL Research University, CNRS, Sorbonne Universit\'es, UPMC Univ. Paris 06, F-92190, Meudon, France}}\\%
$^{d}${\em{School of Earth and Space Exploration, Arizona State University, Tempe, AZ 85281, USA}}\\%
$^{e}${\em{Synchrotron SOLEIL, Orme de Merisiers, St. Aubin, BP 48, F-91192 Gif sur Yvette Cedex, France}}\\%
$^{f}${\em{Department of Chemistry, Massachusetts Institute of Technology, Cambridge, MA 02139, USA}}\\%
}

\date{\today}

\maketitle

\begin{abstract}
  \label{abstract} We have reinvestigated the \As$(v=0)$ level of \CO\ using new high-resolution spectra obtained via multi-photon laser excitation as well as with synchrotron-based Fourier-transform absorption spectroscopy of the \AX$(0, 0)$, \eX$(1, 0)$, \dX$(4, 0)$, \apX$(9, 0)$, and \atX$(11, 0)$ bands.
  In addition, Fourier-transform emission spectroscopy in the visible range is performed on the \BA$(0, 0)$ band.
  Spectra of the \BX$(0, 0)$ band are measured in order to tie information from the latter emission data to the level structure of \As$(v=0)$.
  The high pressures in the absorption cell at the synchrotron and the high temperatures
  in the emission discharge permitted monitoring of high rotational quantum levels in \As$(v=0)$ up to $J=43$.
  All information, in total over 900 spectral lines, was included in an effective-Hamiltonian analysis of the A$^1\Pi(v=0, J)$ levels that are directly perturbed by the \es$(v=1)$, \ds$(v=4)$, \aps$(v=9)$, \Ds$(v=0)$, \Is$(v=0, 1)$ close-lying levels and the \es$(v=0,2)$, \ds$(v=3,5)$, \aps$(v=8,10)$ remote levels, as well being indirectly influenced by the \ats$(v=10, 11)$ state. The influence of nine further perturber levels and their interactions was investigated and are not  significant for reproducing the present experimental data. This analysis leads to a much improved description in terms of molecular constants and interaction parameters, compared to previous studies of the same energy region for other CO isotopologues.

\end{abstract}

\clearpage

\section{Introduction}
\label{sec:Intro}

The spectroscopy of the carbon monoxide molecule is of major
importance in view of its being the second most abundant
molecule in the Universe. Its dipole moment is a decisive
ingredient in the cooling process of interstellar clouds \emph{en route}
to star formation. The probing of CO under a variety of conditions
is crucial to an understanding of the physics and chemistry of the
interstellar medium \cite{Sheffer2007}, of protoplanetary disks
\cite{Perez2015}, of exoplanetary atmospheres \cite{Heng2016},
of galactic structure at large redshifts \cite{Noterdaeme2017}, and
it may turn out to be a probe of temporal variation of fundamental constants
\cite{Dapra2016,Dapra2017a}. In view of saturation and shielding effects of
the strongest transitions, the use and investigation of lower-abundance
isotope-substituted species is of relevance, in particular where
photo-dissociation becomes strongly isotope dependent
\cite{Eidelsberg1990,Eidelsberg1991}, in some cases connected to
subtle effects of perturbations \cite{Cacciani1998,Ubachs2000}.

The CO molecule is a prototypical system for investigating
perturbations in the spectra of diatomic molecules, as is known since
the studies by Field on the \As{} state
\cite{Field1972,Field1972a}. In recent years our team has been
involved in detailed re-investigations of perturbations in the
\As{} state of CO, exploiting a combination of various precision spectroscopic
techniques, where the lowest $v=0$ vibrational level was chosen as
a main target. One of the aims of pursuing a precision study of the \As{} state was the derivation of sensitivity
coefficients for probing a possible variation of fundamental constants
based on the \AX\ system of CO \cite{Salumbides2012}. Thereafter, precision studies of
the \AX$(0, 0)$ bands and the perturbing states were performed for $^{12}$C$^{16}$O \cite{Niu2013},
for $^{13}$C$^{16}$O
\cite{Niu2016b}, for $^{13}$C$^{17}$O \cite{Hakalla2017}, and for
$^{12}$C$^{18}$O \cite{Trivikram2017}. Here we extend these
studies on the \CO{} isotopologue using laser-based excitation and
VUV-Fourier-transform (FT-VUV) absorption spectroscopy as well as
visible Fourier-transform (FT-VIS) emission spectroscopy to observe and assign the perturbations in the
\As($v=0$) state. In the other isotopologues, perturbing effects of
the \aps($v=9$), \ds($v=4$), \es($v=1$), \Is($v=0, 1$), and
\Ds($v=0$) levels were found and these will also be investigated here.
Also, perturbations by levels of the \ats{} state will be addressed.
Therefore, in addition to low-pressure FT-VUV studies performed, focusing on the
\AX\ excitation, high \CO{} pressures were used to observe the weak absorption of the \atX, \dX, \apX, and \eX systems.
These measurements provide additional and accurate information about the perturbing effects on the \As{} state.
In particular, the intensity borrowing effects between the singlet and triplet systems tightly
constrain the values of the perturbation parameters.

Although the \AX{} system of CO has been investigated in many studies over the decades,
the information on the \CO{} isotopologue is scarce. Haridass and coworkers have performed detailed studies of
\As($v=0$) by VUV emission, revealing perturbation effects from the \aps($v=10$) and \ds($v=5$) levels
\cite{Haridass1994b,Haridass1994c}.
Emission studies of the \AA ngstr\"{o}m (\BA) bands of \CO~\cite{Prasad1984,Malak1984} provided further information
on the \As($v=0$) level, and the study of the Herzberg (\CA) systems \cite{Kepa1988,Kepa2014} also provided detailed information
on the interaction with the \es($v=1$) level.
The \AX\ system was reinvestigated recently with the FT-VUV spectrometer at the SOLEIL synchrotron \cite{Lemaire2016},
focusing on the determination of term values and line strength parameters. In that study, the
existence of two additional dipole-allowed singlet systems, denoted as \OSX{} and \OPX, was hypothesised.
Information on the perturbing triplet states in \CO{} has not yet been reported, except for the
study of the \ats($v=0$) state \cite{Denijs2011}.

The present study entails a high-precision re-analysis of the
level structure of the \As($v=0$) state of \CO, following the
rotational manifold up to the rotational quantum number
$J=43$. This results in a much improved description in terms of
molecular constants and interaction parameters, and a comparison
is made with previous studies.

\section{Experimental procedures}
\label{sec:Exp}

As in previous studies on the $^{13}$C$^{16}$O \cite{Niu2016b}
and $^{12}$C$^{18}$O \cite{Trivikram2017}
isotopologues, three distinct experimental techniques are employed to assess
the rotational level structure of the \As($v=0$) manifold. The most
accurate \AX$(0, 0)$ line frequencies were derived from Doppler-free
measurements using a narrowband laser source, consisting of a
pulsed-dye-amplifier (PDA) injection seeded by the continuous-wave
output of a ring-dye laser \cite{Ubachs1997}. A $2+1'$ resonance
enhanced multi-photon ionisation laser scheme was used for
two-photon excitation of the low $J$ rotational levels of the \AX$(0,0)$
band, followed by ionisation by a second UV laser pulse at 203 nm,
as described in Niu et al. \cite{Niu2015}. Mass-dependent detection of CO
isotopologues was achieved in a time-of-flight mass spectrometer.
An isotopically enriched \CO{} gas sample was used for the
experiment (Sigma Aldrich, 99\% atom $^{13}$C and 95\% atom
$^{18}$O). Absolute frequencies were determined by simultaneous
recording of saturated I$_2$ resonances \cite{Xu2000} and markers
from a stabilised etalon. Frequency chirp effects in the PDA laser
were measured and corrected for off-line, but their uncertainties nevertheless contribute decisively
to the uncertainty budget of the laser experiments. The transition frequencies were
measured for a range of laser intensities in the focal region to
assess the AC-Stark effect; extrapolation to zero intensity
resulted in the true transition frequencies. The overall accuracy
of the transition frequencies falls between $0.002-0.003$ \wn. The
laser-based experiments were carried out at LaserLaB Amsterdam.

Visible emission data were recorded from a hollow-cathode
discharge lamp. It was initially filled with a mixture of helium
and acetylene $^{13}$C$_2$D$_2$ (Cambridge Isotopes, 99.98\% of
$^{13}$C) at a pressure of approximately 10 mbar. A DC electric
current was passed through the mixture for about 150 hours. The
process, similar to that described by Hakalla et al. \cite{Hakalla2013,
Hakalla2017}, resulted in the deposition of a small amount of $^{13}$C
inside the cathode. Next, the lamp was evacuated and filled with
an enriched sample of $^{18}$O$_2$ gas (Sigma-Aldrich, 98.1\%).
The electrodes were operated at 950\,V and 80\,mA DC with a static gas pressure of 3 mbar. The higher
temperature of the \CO{} plasma formed at the center of the cathode,
up to 1000 K, allowed for observations of rotational
transitions with $J$ up to 41; a higher value than in our previous experiments \cite{Hakalla2012c, Hakalla2012b}.
The physical line-broadening increased by only $0.02$ \wn\ relative to that reported in
\cite{Hakalla2014,Hakalla2014b}, where the temperatures of plasmas
were about 650\ K.  The final molecular gas composition used to
obtain the spectrum was \CO{} : $^{12}$C$^{18}$O $= 1 : 0.1$.
The spectral emission was analysed by a 1.7 m Fourier-transform
(FT) spectrometer used under vacuum conditions. Operation of the
setup and calibration procedures were explained in recent reports
on the setup installed at Rzesz\'{o}w University
\cite{Niu2016b,Hakalla2017,Trivikram2017}. Spectra were
accumulated over $128$ scans with a spectral resolution of $0.018$ \wn. The accuracy
on the transition frequencies amounts to $0.003-0.03$ \wn.

The setup involving the Vacuum UltraViolet Fourier-transform  (FT-VUV) spectrometer at the SOLEIL
synchrotron \cite{Deoliveira2011,Deoliveira2016} was employed to obtain absorption spectra of \CO{} under three regimes.
First, spectra of the \AX$(0,0)$ band were measured under conditions of low gas density in
a quasi-static flow through a windowless cell, with column densities in the range \np{e14} to \np[cm^{-2}]{2e15}, as in previous
studies \cite{Niu2013,Niu2016,Gavilan2013,Lemaire2016}.
This provides spectroscopic information on $J$-levels up to about 20.
Second, further \AX$(0,0)$ spectra were recorded at high gas
pressures (up to about 80\,mbar) in a closed cell of 9\,cm length and sealed by magnesium-fluoride windows \cite{Gavilan2013}.
This enabled probing lines with $J$ up to 43 and the detection of many more lines of the perturbing  \eX, \dX, \apX, and \atX{} forbidden band systems. An approximate column density of \np[cm^{-2}]{2e19} was achieved in this case.
The full-width half-maximum self-broadening of lines in the \AX$(20,0)$ and $(21,0)$ bands of ${}^{12}$C$^{16}$O is measured to be $(2.3\pm 0.5)\times 10^{-4}$\,cm$^{-1}$\,mbar$^{-1}$ \cite{stark1998}, leading to a maximum broadening in our case of 0.02\,cm$^{-1}$ that is not detectable due to significantly-greater broadening arising from the Doppler effect and finite instrumental resolution.

Finally, the \BX$(0, 0)$ band was recorded in a heated windowless cell, attaining a rotational temperature of $\sim 1000$\,K in a setup similar to that of Niu et al. \cite{Niu2015a}.
Transition frequencies deduced from FT-VUV spectra have accuracies estimated to fall in the range $0.02 - 0.05$ \wn,
depending on the specific conditions under which the spectra were recorded and the blendedness, weakness, or saturation of individual absorption lines. The lower uncertainty limit generally applies to the low-density room-temperature spectra, while greater uncertainties are associated with higher temperature and pressure spectra.

\section{Results}
\label{sec:Res}

 In the following the results of all individual studies are presented.

 \subsection{Results from laser-based study}

\begin{figure}
  \centering
  \resizebox{0.6\textwidth}{!}{\includegraphics{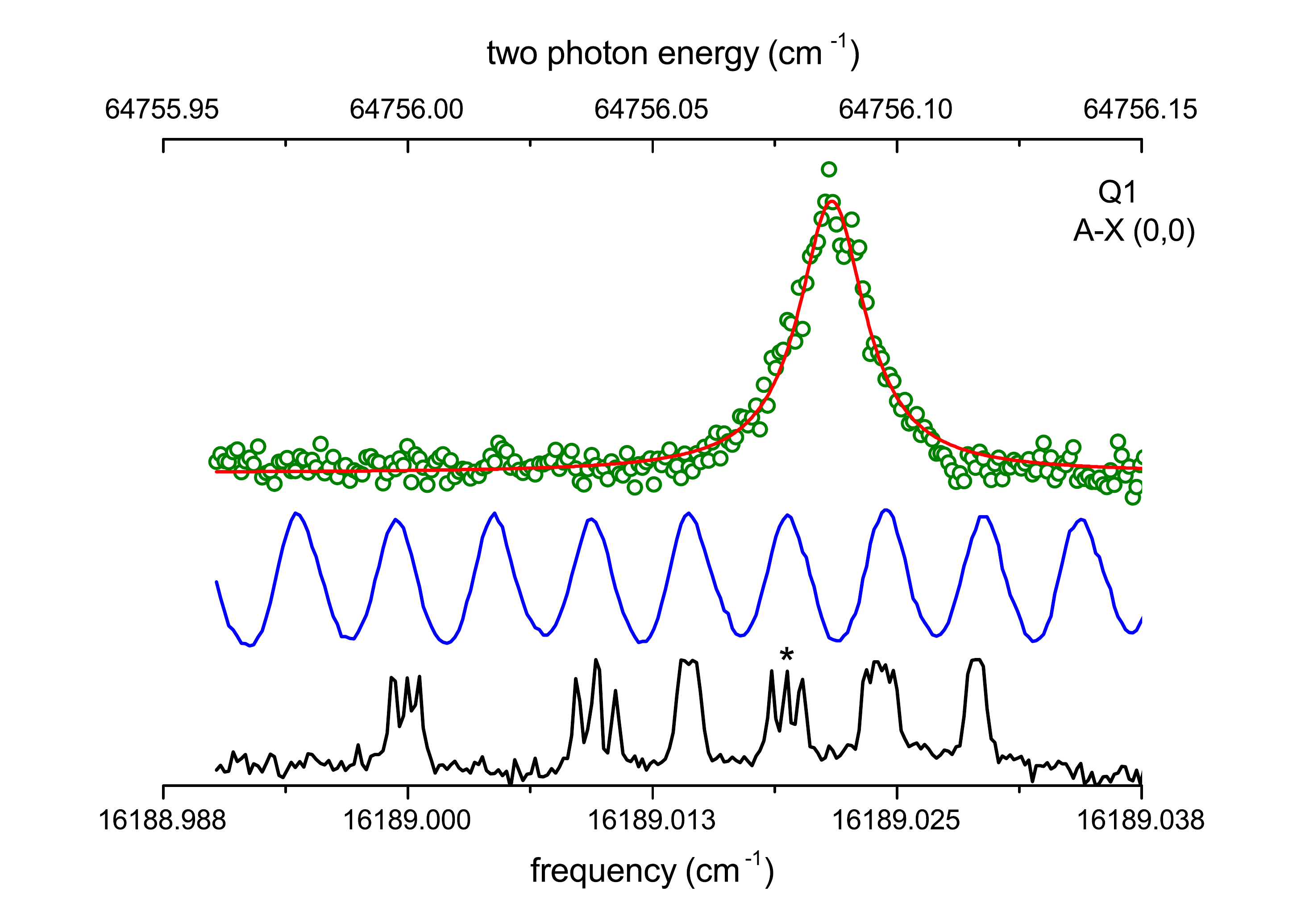}}
  \caption{The \AX $(0,0)$ $Q$(1) transition measured by $2+1'$ REMPI (open green points in upper curve) and fitted (red curve). The middle blue line and lower black line show etalon markers from a stabilised Fabry-Perot interferometer and the saturated iodine spectrum used for frequency interpolation and calibration, respectively. The asterisk indicates the a13 hyperfine component of the B-X $(10,3)$ $R$(87) iodine line at $16\,189.019\,45$ \wn~\cite{Xu2000} that was used for an absolute calibration. }
  \label{PDAspec}
\end{figure}

\begin{figure}
  \centering
\resizebox{0.48\textwidth}{!}{\includegraphics{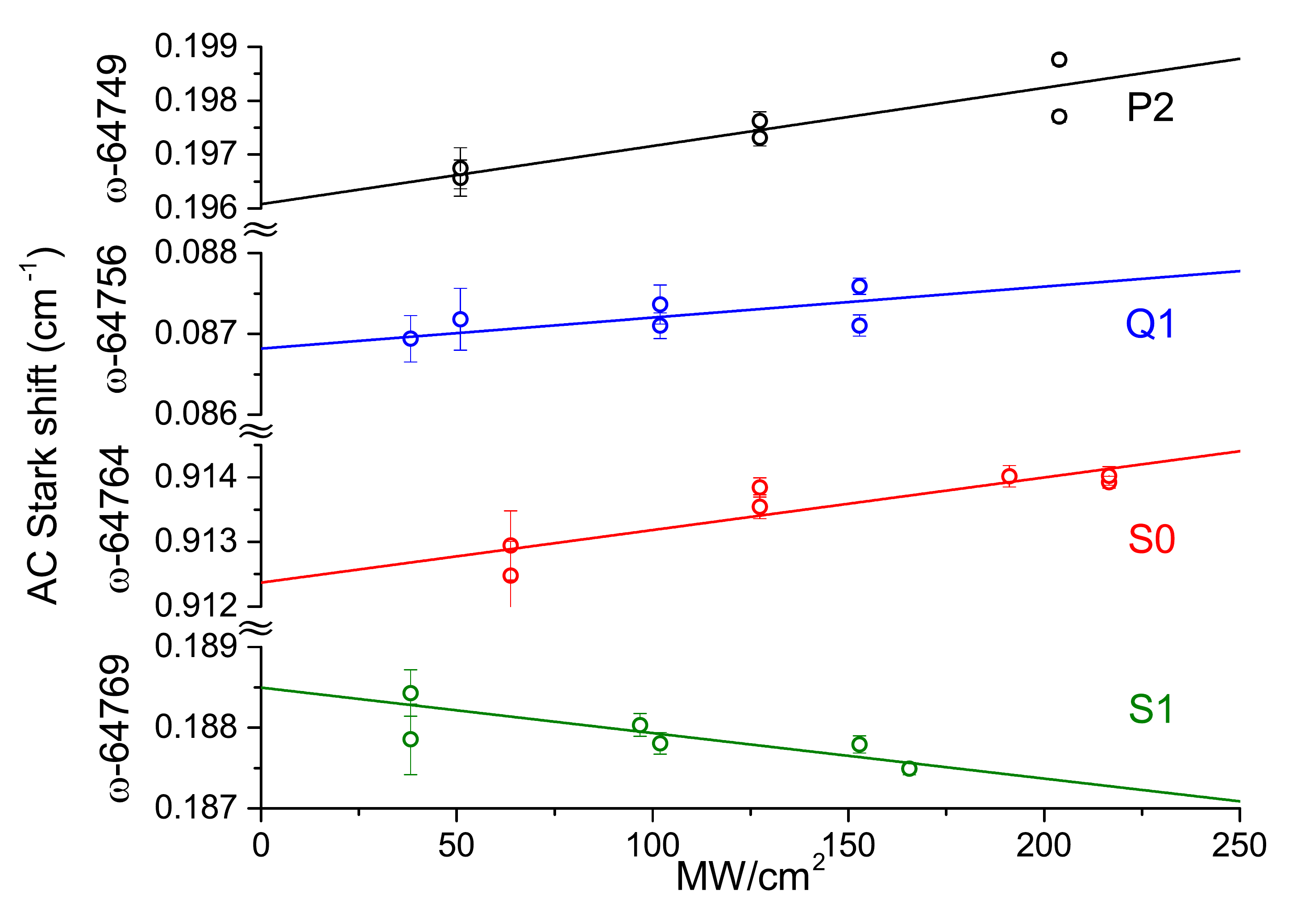}}
\caption{AC-Stark-plots for four \AX$(0,0)$ two-photon transitions.}
\label{AC-stark}
\end{figure}

Laser-based $2+1'$ REMPI spectra were recorded for nine two-photon
transitions of the \AX$(0, 0)$ band. Figure \ref{PDAspec} displays a
spectrum of the $Q$(1) line. All line measurements were
performed as a function of laser power density in the focal region,
similarly to the previous study of $^{12}$C$^{18}$O
\cite{Trivikram2017}. Figure \ref{AC-stark} shows the AC-Stark
extrapolation curves for four selected lines. The transition frequencies were derived
from extrapolation to zero-intensity levels and are listed in Table \ref{Laser-list}.

Interestingly, the sign of the AC-Stark slope is negative for the $S$(1) two-photon transition, while the slope
for the other transitions is positive (see Fig.~\ref{AC-stark}).
From a reanalysis of the data reported in \cite{Niu2016b} it is found that
the sign of the AC-Stark slope for the $S$(1) line in $^{13}$C$^{16}$O is also negative,
while all other lines exhibit a positive AC-Stark slope. In contrast, for all lines in
$^{12}$C$^{18}$O \cite{Trivikram2017} a positive AC-Stark slope is found.
This phenomenon is connected to the molecular level structure
at the three-photon excitation level in the molecule and can be studied in further detail
by performing two-color ionisation experiments \cite{Xu1994}.

\begin{table}
  \centering
  \caption{Results from the two-photon Doppler-free laser experiment. Measured \AX $(0,0)$ transition frequencies, $\nu_{\rm{obs}}$, and AC-Stark slope coefficients, $C_{\rm{AC}}$, for four selected lines. }
  \label{Laser-list}
  \small
  \begin{threeparttable}
  \begin{tabular}
    {l@{\hspace{15pt}}cc}
    \colrule
    \multicolumn{1}{l}{Line} &
    \multicolumn{1}{c}{$\nu_{\rm{obs}}$\textsuperscript{$a$}} &
    \multicolumn{1}{r}{$C_{\rm{AC}}$\textsuperscript{$b$}}  \\
    \colrule
    $P$(2)    &   64\,749.1956\,(20)  &  0.32 \\
    $P$(3)    &   64\,744.3124\,(30)  &  \\
    $Q$(1)    &   64\,756.0869\,(30)  &  0.11 \\
    $Q$(2)    &   64\,754.4352\,(20)  &   \\
    $R$(1)    &   64\,761.7759\,(30)  &   \\
    $R$(2)    &   64\,763.1306\,(20)  &   \\
    $S$(0)    &   64\,764.9124\,(30)  &  0.24 \\
    $S$(1)    &   64\,769.1883\,(30)  &  $-$0.16 \\
    $S$(2)    &   64\,771.9897\,(30)  &   \\
    \colrule
  \end{tabular}
  \begin{tablenotes}
    \footnotesize
    \item[$a$] Units of \wn{} and $1\sigma$ uncertainties given in parentheses in units of the least-significant digit.
    \item[$b$] Units of MHz/(MW/cm$^2$).
  \end{tablenotes}
  \end{threeparttable}
\end{table}

\subsection{Results from FT-VIS study}

Here we present the results of the Fourier-transform emission study of the \CO\ \BA$(0, 0)$ band and the perturbing lines in the wavelength range $\lambda=438-452$ nm, with a measured spectrum shown in Fig.~\ref{fig:BA00} (an electronic form of the Fig.~\ref{fig:BA00}{} source spectrum is given in the Supplementary Material). This range covers an interval where perturber-state lines of \Be$(0, 1)$, \Bd$(0, 4)$, and \Bap$(0, 9)$ bands can be identified.
Source contamination by $^{12}$C$^{18}$O \BA$(0, 0)$, \CO\ \BA$(1, 1)$ and $^{12}$C$^{18}$O \BA$(1, 1)$ bands, was taken into consideration during the analysis. Measured transition frequencies of \CO\ \BA$(0, 0)$ are listed in Table \ref{VIS-FTS lines}, while the lines connecting the \Bs($v=0$) upper state to perturber levels are listed in Table \ref{VIS-FTS extra-lines}.
Line positions were measured by fitting Voigt lineshape functions to the experimental lines. Line position uncertainties were evaluated using an empirical relation similar to that given by Brault \cite{Brault1987}:
\begin{equation}\label{eq1}
\Delta \sigma = \frac{f}{\sqrt{N}}\frac{FWHM}{\sqrt{SNR}},
\end{equation}
where $f$ is a constant of the order of unity that is lineshape dependent, $FWHM$ is the full-width at half-maximum of the line, $N$ is the true number of statistically independent points in a linewidth (taking into account the zero filling factor commonly used to interpolate FT spectra), and $SNR$ is the signal-to-noise ratio.

\begin{figure}
\begin{center}
\includegraphics[width=\textwidth]{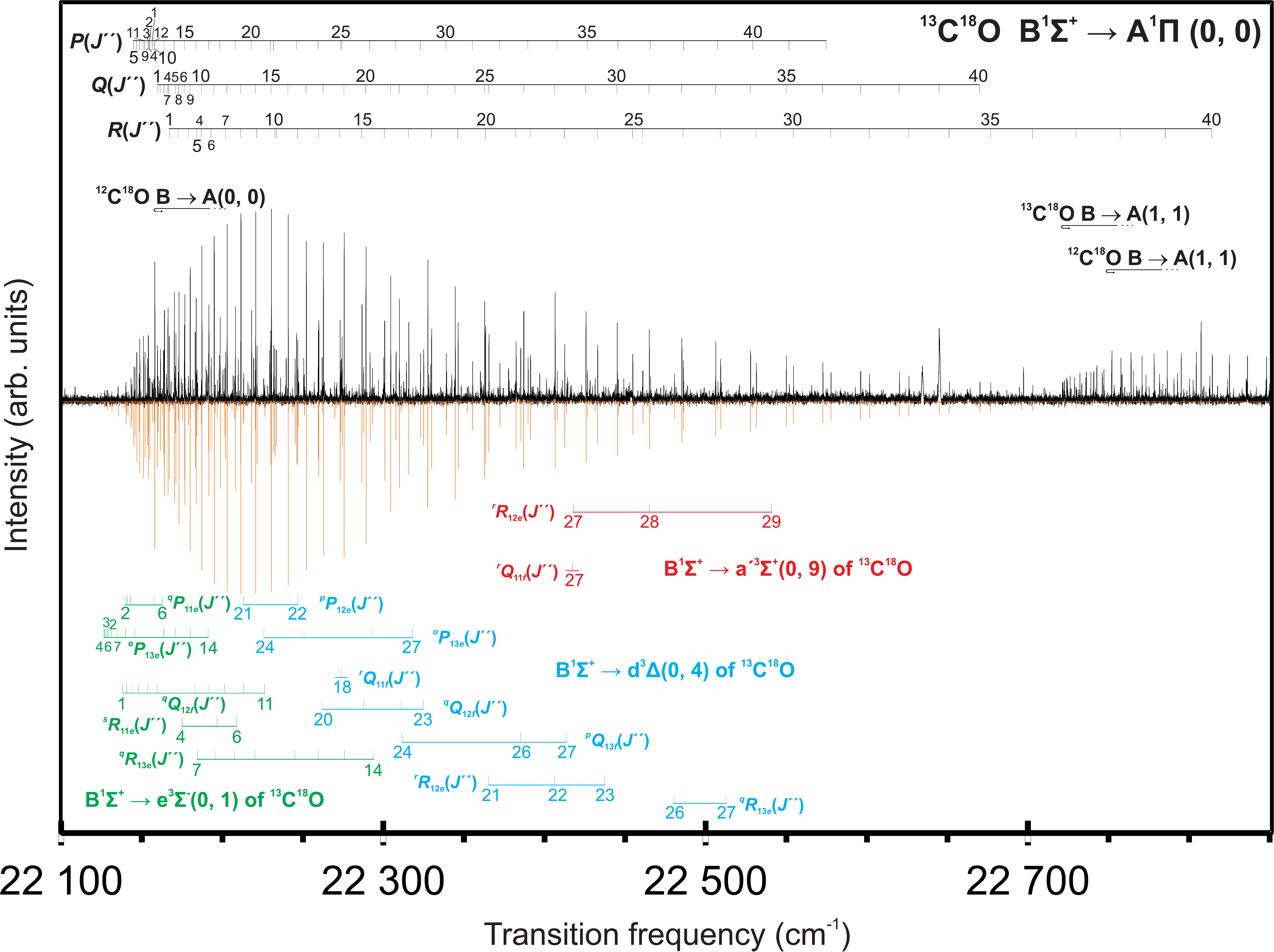}
\caption{VIS high-resolution photoemission spectra recorded by the FTS technique. Observed \CO{} rovibronic bands are indicated. The upper trace presents an experimental spectrum, whereas the lower one shows a simulated spectrum of the \CO \ \BA $(0, 0)$ band and lines terminating on states perturbing \As(\0). The simulation is obtained using the PGOPHER software \cite{pgopher}.}
\label{fig:BA00}
\end{center}
\end{figure}

\begin{table} [h]
  \centering
  \caption{Transition frequencies of the \CO \  \BA \ $(0,0)$ band obtained from the FT-VIS experiment.$^{a}$}
  \label{VIS-FTS lines}
  \footnotesize
  \begin{threeparttable}
    \begin{tabular}
      {l@{\hspace{20pt}}l@{\hspace{3pt}}r@{\hspace{3pt}}l@{\hspace{20pt}}l@{\hspace{3pt}}r@{\hspace{3pt}}l@{\hspace{20pt}}l@{\hspace{3pt}}r@{\hspace{3pt}}l}

\colrule																																						
$J''$	&		$R(J'')$				$^{		}$	&	$o-c$	&		&		$Q(J'')$				$^{		}$	&	$o-c$	&		&		$P(J'')$				$^{		}$	&	$o-c$	\\
\colrule	&						$^{		}$	&		&		&						$^{		}$	&		&		&						$^{		}$	&		\\
1	&	22	168.40	(	2	)	$^{	bw	}$	&	0.01	&		&	22	161.219	(	6	)	$^{	b	}$	&	-0.005	&		&	22	157.80	(	2	)	$^{	bw	}$	&	0.02	\\
2	&	22	173.685	(	8	)	$^{	bw	}$	&	0.003	&		&	22	162.736	(	6	)	$^{	b	}$	&	0.025	&		&	22	155.992	(	6	)	$^{	bw	}$	&	0.007	\\
3	&	22	180.077	(	6	)	$^{	b	}$	&	0.005	&		&	22	164.996	(	3	)	$^{	b	}$	&	0.009	&		&	22	155.295	(	5	)	$^{		}$	&	-0.003	\\
4	&	22	187.995	(	7	)	$^{	b	}$	&	0.019	&		&	22	168.127	(	2	)	$^{		}$	&	0.007	&		&	22	156.125	(	6	)	$^{	bw	}$	&	0.001	\\
5	&	22	185.288	(	5	)	$^{	b	}$	&	0.003	&		&	22	172.235	(	2	)	$^{		}$	&	0.007	&		&	22	146.351	(	4	)	$^{	b	}$	&	-0.008	\\
6	&	22	194.047	(	4	)	$^{		}$	&	-0.001	&		&	22	177.491	(	2	)	$^{		}$	&	-0.007	&		&	22	148.048	(	3	)	$^{		}$	&	0.002	\\
7	&	22	202.956	(	3	)	$^{		}$	&	-0.002	&		&	22	167.340	(	2	)	$^{		}$	&	-0.003	&		&	22	149.885	(	3	)	$^{		}$	&	0.002	\\
8	&	22	212.326	(	3	)	$^{	b	}$	&	-0.002	&		&	22	174.118	(	2	)	$^{		}$	&	-0.001	&		&	22	152.182	(	2	)	$^{		}$	&	-0.001	\\
9	&	22	222.401	(	3	)	$^{		}$	&	-0.004	&		&	22	181.087	(	1	)	$^{		}$	&	-0.002	&		&	22	155.188	(	2	)	$^{		}$	&	-0.004	\\
10	&	22	233.471	(	3	)	$^{		}$	&	-0.004	&		&	22	188.309	(	1	)	$^{		}$	&	-0.002	&		&	22	159.191	(	1	)	$^{	b	}$	&	-0.006	\\
11	&	22	234.468	(	3	)	$^{		}$	&	0.010	&		&	22	195.890	(	1	)	$^{		}$	&	-0.002	&		&	22	153.110	(	4	)	$^{	b	}$	&	-0.008	\\
12	&	22	247.593	(	3	)	$^{		}$	&	0.007	&		&	22	203.925	(	1	)	$^{		}$	&	-0.001	&		&	22	159.191	(	1	)	$^{	b	}$	&	0.006	\\
13	&	22	260.615	(	2	)	$^{		}$	&	0.006	&		&	22	212.475	(	1	)	$^{		}$	&	-0.002	&		&	22	165.159	(	2	)	$^{		}$	&	0.007	\\
14	&	22	273.776	(	2	)	$^{		}$	&	0.005	&		&	22	221.591	(	1	)	$^{		}$	&	-0.002	&		&	22	171.269	(	2	)	$^{	b	}$	&	0.009	\\
15	&	22	287.312	(	2	)	$^{	b	}$	&	0.005	&		&	22	231.313	(	1	)	$^{		}$	&	-0.002	&		&	22	177.753	(	2	)	$^{		}$	&	0.006	\\
16	&	22	301.344	(	2	)	$^{	*	}$	&	-0.037	&		&	22	241.693	(	1	)	$^{		}$	&	-0.002	&		&	22	184.781	(	2	)	$^{		}$	&	0.005	\\
17	&	22	316.275	(	2	)	$^{		}$	&	0.001	&		&	22	252.908	(	1	)	$^{	*	}$	&	-0.045	&		&	22	192.633	(	2	)	$^{	*	}$	&	0.004	\\
18	&	22	330.357	(	3	)	$^{		}$	&	0.008	&		&	22	263.422	(	1	)	$^{		}$	&	-0.005	&		&	22	199.627	(	2	)	$^{	*	}$	&	-0.041	\\
19	&	22	346.758	(	2	)	$^{		}$	&	0.007	&		&	22	276.290	(	1	)	$^{		}$	&	0.002	&		&	22	209.038	(	2	)	$^{		}$	&	-0.001	\\
20	&	22	363.791	(	2	)	$^{		}$	&	0.003	&		&	22	289.781	(	1	)	$^{		}$	&	-0.001	&		&	22	219.056	(	2	)	$^{		}$	&	0.006	\\
21	&	22	382.617	(	3	)	$^{	b	}$	&	0.002	&		&	22	305.061	(	1	)	$^{		}$	&	-0.005	&		&	22	230.859	(	2	)	$^{		}$	&	0.003	\\
22	&	22	391.537	(	3	)	$^{		}$	&	0.004	&		&	22	310.448	(	2	)	$^{		}$	&	0.004	&		&	22	232.767	(	3	)	$^{		}$	&	0.007	\\
23	&	22	412.671	(	3	)	$^{		}$	&	0.008	&		&	22	328.087	(	1	)	$^{		}$	&	0.001	&		&	22	246.892	(	3	)	$^{		}$	&	0.011	\\
24	&	22	433.056	(	3	)	$^{		}$	&	0.009	&		&	22	344.976	(	2	)	$^{		}$	&	0.004	&		&	22	260.259	(	3	)	$^{	b	}$	&	-0.004	\\
25	&	22	454.772	(	3	)	$^{	b	}$	&	0.010	&		&	22	363.200	(	2	)	$^{		}$	&	0.002	&		&	22	274.991	(	3	)	$^{	b	}$	&	0.009	\\
26	&	22	460.981	(	5	)	$^{		}$	&	-0.010	&		&	22	365.815	(	2	)	$^{	b	}$	&	-0.009	&		&	22	274.202	(	4	)	$^{	b	}$	&	-0.020	\\
27	&	22	486.036	(	3	)	$^{		}$	&	-0.004	&		&	22	387.367	(	2	)	$^{		}$	&	-0.001	&		&	22	292.283	(	3	)	$^{		}$	&	-0.004	\\
28	&	22	509.103	(	4	)	$^{		}$	&	-0.003	&		&	22	406.747	(	2	)	$^{		}$	&	0.001	&		&	22	308.379	(	3	)	$^{		}$	&	0.001	\\
29	&	22	531.185	(	4	)	$^{	b	}$	&	-0.009	&		&	22	425.817	(	2	)	$^{		}$	&	0.004	&		&	22	323.505	(	3	)	$^{		}$	&	0.006	\\
30	&	22	554.083	(	4	)	$^{	b	}$	&	-0.007	&		&	22	445.157	(	2	)	$^{		}$	&	0.003	&		&	22	339.437	(	3	)	$^{		}$	&	0.001	\\
31	&	22	577.379	(	6	)	$^{	b	}$	&	0.013	&		&	22	465.003	(	2	)	$^{		}$	&	-0.004	&		&	22	355.751	(	4	)	$^{	b	}$	&	-0.011	\\
32	&	22	601.178	(	7	)	$^{	b	}$	&	0.015	&		&	22	485.099	(	2	)	$^{		}$	&	-0.013	&		&	22	372.614	(	4	)	$^{	b	}$	&	-0.004	\\
33	&	22	625.517	(	7	)	$^{	b	}$	&	-0.012	&		&	22	506.080	(	3	)	$^{		}$	&	0.001	&		&	22	390.021	(	3	)	$^{	b	}$	&	-0.027	\\
34	&	22	650.496	(	8	)	$^{	*	}$	&	0.004	&		&	22	527.581	(	3	)	$^{		}$	&	-0.011	&		&	22	408.071	(	5	)	$^{	b	}$	&	-0.015	\\
35	&	22	676.090	(	8	)	$^{	*b	}$	&	-0.044	&		&	22	549.742	(	3	)	$^{	*b	}$	&	-0.011	&		&	22	426.779	(	6	)	$^{	*b	}$	&	-0.033	\\
36	&	22	702.125	(	8	)	$^{	b	}$	&	0.068	&		&	22	572.317	(	4	)	$^{	*b	}$	&	0.048	&		&	22	445.881	(	6	)	$^{	*b	}$	&	0.054	\\
37	&	22	728.91	(	2	)	$^{	bw	}$	&	0.03	&		&	22	595.689	(	6	)	$^{	b	}$	&	0.028	&		&	22	465.779	(	6	)	$^{	b	}$	&	0.018	\\
38	&	22	756.299	(	9	)	$^{	bw	}$	&	0.007	&		&	22	619.635	(	5	)	$^{	b	}$	&	0.007	&		&	22	486.292	(	7	)	$^{	b	}$	&	0.013	\\
39	&	22	784.31	(	1	)	$^{	bw	}$	&	0.01	&		&	22	644.228	(	2	)	$^{	b	}$	&	0.012	&		&	22	507.427	(	7	)	$^{	b	}$	&	0.005	\\
40	&	22	812.93	(	3	)	$^{	bw	}$	&	0.03	&		&	22	669.458	(	7	)	$^{	bw	}$	&	0.027	&		&	22	529.184	(	9	)	$^{	bs	}$	&	-0.014	\\
41	&									&		&		&									&		&		&	22	551.57	(	3	)	$^{	bbs	}$	&	-0.04	\\
42	&									&		&		&									&		&		&	22	574.62	(	3	)	$^{	bbs	}$	&	-0.05	\\
\colrule																																						
																																						
    \end{tabular}
    \begin{tablenotes}
      \footnotesize
      \item[$a$] In units of \wn. The number in parentheses indicates the uncertainty of expected line position given by the empirical relation in Eq.~(\ref{eq1}). Lines marked with $b$ and/or $w$ are blended and/or weak. For lines marked by an asterisk (*) a very weak perturbation in the \Bs($v=0$) Rydberg state was found. The instrumental resolution was 0.018 \wn. The estimated absolute calibration uncertainty (1$\sigma$) was 0.003 \wn. The column with $o-c$ displays the deviations between observed values and values calculated by the fitting routine.
    \end{tablenotes}
  \end{threeparttable}
\end{table}

\begin{table}
  \centering
  \caption{Spin-forbidden lines appearing in the FT-VIS emission spectrum of \CO.$^{a}$}
  \label{VIS-FTS extra-lines} \footnotesize
  \footnotesize
  \begin{threeparttable}
    \begin{tabular}
      {l@{\hspace{20pt}}l@{\hspace{3pt}}r@{\hspace{3pt}}l@{\hspace{20pt}}l@{\hspace{3pt}}r@{\hspace{3pt}}l@{\hspace{20pt}}l@{\hspace{3pt}}r@{\hspace{3pt}}l}
 \colrule																																						
$J''$	&		$^sR_{11ee}$				$^{		}$	&	$o-c$	&		&		$^rQ_{11ef}$				$^{		}$	&	$o-c$	&		&		$^qP_{11ee}$				$^{		}$	&	$o-c$	\\
 \colrule	&						$^{		}$	&		&		&						$^{		}$	&		&		&						$^{		}$	&		\\
	&						$^{		}$	&		&		&		\Be$(0, 1)$				$^{		}$	&		&		&						$^{		}$	&		\\
2	&						$^{		}$	&		&		&						$^{		}$	&		&		&	22	141.04	(	2	)	$^{	bw	}$	&	-0.02	\\
3	&						$^{		}$	&		&		&						$^{		}$	&		&		&	22	142.58	(	2	)	$^{	bw	}$	&	0.02	\\
4	&	22	176.344	(	8	)	$^{	b	}$	&	-0.011	&		&						$^{		}$	&		&		&	22	144.496	(	6	)	$^{	b	}$	&	-0.008	\\
5	&	22	197.886	(	6	)	$^{	b	}$	&	-0.004	&		&						$^{		}$	&		&		&	22	158.968	(	6	)	$^{		}$	&	0.005	\\
6	&	22	209.90	(	2	)	$^{	b	}$	&	-0.02	&		&						$^{		}$	&		&		&	22	163.92	(	2	)	$^{	bw	}$	&	0.01	\\
	&						$^{		}$	&		&		&		\Bd$(0, 4)$				$^{		}$	&		&		&						$^{		}$	&		\\
	&						$^{		}$	&		&		&						$^{		}$	&		&		&						$^{		}$	&		\\
18	&						$^{		}$	&		&		&	22	270.377	(	8	)	$^{		}$	&	0.006	&		&						$^{		}$	&		\\
19	&						$^{		}$	&		&		&						$^{		}$	&		&		&						$^{		}$	&		\\
	&						$^{		}$	&		&		&		\Bap$(0, 9)$				$^{		}$	&		&		&						$^{		}$	&		\\
	&						$^{		}$	&		&		&						$^{		}$	&		&		&						$^{		}$	&		\\
27	&						$^{		}$	&		&		&	22	418.12	(	3	)	$^{	bw	}$	&	0.02	&		&						$^{		}$	&		\\
 \colrule	&						$^{		}$	&		&		&						$^{		}$	&		&		&						$^{		}$	&		\\
	&		$^rR_{12ee}$				$^{		}$	&	$o-c$	&		&		$^qQ_{12ef}$				$^{		}$	&	$o-c$	&		&		$^pP_{12ee}$				$^{		}$	&	$o-c$	\\
 \colrule	&						$^{		}$	&		&		&						$^{		}$	&		&		&						$^{		}$	&		\\
	&						$^{		}$	&		&		&		\Be$(0, 1)$				$^{		}$	&		&		&						$^{		}$	&		\\
	&						$^{		}$	&		&		&						$^{		}$	&		&		&						$^{		}$	&		\\
1	&						$^{		}$	&		&		&	22	139.49	(	2	)	$^{	bw	}$	&	0.04	&		&						$^{		}$	&		\\
2	&						$^{		}$	&		&		&	22	141.72	(	2	)	$^{	bw	}$	&	-0.02	&		&						$^{		}$	&		\\
3	&						$^{		}$	&		&		&	22	145.115	(	9	)	$^{	b	}$	&	0.010	&		&						$^{		}$	&		\\
4	&						$^{		}$	&		&		&	22	149.509	(	7	)	$^{	b	}$	&	0.008	&		&						$^{		}$	&		\\
5	&						$^{		}$	&		&		&	22	154.822	(	4	)	$^{	b	}$	&	0.017	&		&						$^{		}$	&		\\
6	&						$^{		}$	&		&		&	22	160.830	(	2	)	$^{		}$	&	-0.004	&		&						$^{		}$	&		\\
7	&						$^{		}$	&		&		&	22	184.176	(	2	)	$^{		}$	&	0.007	&		&						$^{		}$	&		\\
8	&						$^{		}$	&		&		&	22	192.455	(	3	)	$^{		}$	&	-0.003	&		&						$^{		}$	&		\\
9	&						$^{		}$	&		&		&	22	202.439	(	4	)	$^{	b	}$	&	-0.001	&		&						$^{		}$	&		\\
10	&						$^{		}$	&		&		&	22	214.06	(	2	)	$^{	bw	}$	&	0.01	&		&						$^{		}$	&		\\
11	&						$^{		}$	&		&		&	22	227.202	(	8	)	$^{	bw	}$	&	-0.002	&		&						$^{		}$	&		\\
	&						$^{		}$	&		&		&						$^{		}$	&		&		&						$^{		}$	&		\\
	&						$^{		}$	&		&		&		\Bd$(0, 4)$				$^{		}$	&		&		&						$^{		}$	&		\\
	&						$^{		}$	&		&		&						$^{		}$	&		&		&						$^{		}$	&		\\
20	&						$^{		}$	&		&		&	22	263.02	(	2	)	$^{	b	}$	&	-0.04	&		&						$^{		}$	&		\\
21	&	22	365.93	(	1	)	$^{	b	}$	&	0.02	&		&	22	288.291	(	4	)	$^{	b	}$	&	-0.009	&		&	22	214.153	(	4	)	$^{	b	}$	&	-0.006	\\
22	&	22	406.548	(	5	)	$^{		}$	&	-0.009	&		&	22	325.445	(	3	)	$^{		}$	&	-0.008	&		&	22	247.777	(	4	)	$^{		}$	&	-0.007	\\
23	&	22	437.07	(	2	)	$^{	b	}$	&	-0.01	&		&	22	352.439	(	9	)	$^{	b	}$	&	-0.003	&		&						$^{		}$	&		\\
	&						$^{		}$	&		&		&						$^{		}$	&		&		&						$^{		}$	&		\\
	&						$^{		}$	&		&		&		\Bap$(0, 9)$				$^{		}$	&		&		&						$^{		}$	&		\\
27	&	22	459.58	(	2	)	$^{	b	}$	&	-0.01	&		&						$^{		}$	&		&		&						$^{		}$	&		\\
28	&	22	501.46	(	7	)	$^{	b	}$	&	-0.01	&		&						$^{		}$	&		&		&						$^{		}$	&		\\
29	&	22	545.39	(	2	)	$^{	bw	}$	&	0.02	&		&						$^{		}$	&		&		&						$^{		}$	&		\\
 \colrule	&						$^{		}$	&		&		&						$^{		}$	&		&		&						$^{		}$	&		\\
	&		$^qR_{13ee}$				$^{		}$	&	$o-c$	&		&		$^pQ_{13ef}$				$^{		}$	&	$o-c$	&		&		$^oP_{13ee}$				$^{		}$	&	$o-c$	\\
 \colrule	&						$^{		}$	&		&		&						$^{		}$	&		&		&						$^{		}$	&		\\
	&						$^{		}$	&		&		&		\Be$(0, 1)$				$^{		}$	&		&		&						$^{		}$	&		\\
2	&						$^{		}$	&		&		&						$^{		}$	&		&		&	22	130.75	(	2	)	$^{	bw	}$	&	-0.02	\\
3	&						$^{		}$	&		&		&						$^{		}$	&		&		&	22	128.752	(	4	)	$^{	bw	}$	&	0.022	\\
4	&						$^{		}$	&		&		&						$^{		}$	&		&		&	22	127.921	(	7	)	$^{	bw	}$	&	0.019	\\
5	&						$^{		}$	&		&		&						$^{		}$	&		&		&	22	128.281	(	9	)	$^{	bw	}$	&	0.002	\\
6	&						$^{		}$	&		&		&						$^{		}$	&		&		&	22	129.867	(	4	)	$^{	bw	}$	&	0.016	\\
7	&	22	185.663	(	7	)	$^{	b	}$	&	-0.006	&		&						$^{		}$	&		&		&	22	132.60	(	2	)	$^{	bw	}$	&	0.01	\\
8	&	22	196.597	(	8	)	$^{	b	}$	&	-0.006	&		&						$^{		}$	&		&		&	22	136.43	(	1	)	$^{	bw	}$	&	-0.03	\\
9	&	22	208.55	(	1	)	$^{	b	}$	&	-0.01	&		&						$^{		}$	&		&		&	22	141.345	(	9	)	$^{	b	}$	&	0.005	\\
10	&	22	221.308	(	4	)	$^{	b	}$	&	0.005	&		&						$^{		}$	&		&		&	22	147.028	(	4	)	$^{	b	}$	&	0.004	\\
11	&	22	245.970	(	4	)	$^{	b	}$	&	-0.001	&		&						$^{		}$	&		&		&	22	164.628	(	2	)	$^{		}$	&	-0.003	\\
12	&	22	260.39	(	2	)	$^{	b	}$	&	0.04	&		&						$^{		}$	&		&		&	22	171.949	(	6	)	$^{	b	}$	&	0.002	\\
13	&	22	276.74	(	2	)	$^{	b	}$	&	0.04	&		&						$^{		}$	&		&		&	22	181.231	(	6	)	$^{	b	}$	&	-0.013	\\
14	&	22	294.78	(	2	)	$^{	b	}$	&	-0.03	&		&						$^{		}$	&		&		&	22	192.28	(	2	)	$^{	b	}$	&	-0.01	\\
	&						$^{		}$	&		&		&		\Bd$(0, 4)$				$^{		}$	&		&		&						$^{		}$	&		\\
	&						$^{		}$	&		&		&						$^{		}$	&		&		&						$^{		}$	&		\\
24	&						$^{		}$	&		&		&	22	311.53	(	2	)	$^{	b	}$	&	-0.02	&		&	22	226.91	(	2	)	$^{	b	}$	&	0.01	\\
25	&						$^{		}$	&		&		&						$^{		}$	&		&		&	22	251.703	(	9	)	$^{	b	}$	&	-0.008	\\
26	&	22	480.494	(	6	)	$^{	b	}$	&	-0.013	&		&	22	385.490	(	5	)	$^{	b	}$	&	-0.004	&		&	22	293.740	(	5	)	$^{	b	}$	&	0.002	\\
27	&	22	512.52	(	2	)	$^{		}$	&	-0.02	&		&	22	413.852	(	9	)	$^{	b	}$	&	0.015	&		&	22	318.775	(	9	)	$^{	b	}$	&	-0.013	\\
 \colrule																																						

    \end{tabular}
    \begin{tablenotes}
      \footnotesize
      \item [$a$] {All transition frequencies are in \wn. The number in parentheses indicates the uncertainty of expected line position given by the empirical relation (\ref{eq1}). Lines marked with $w$ and/or $b$ are weak and/or blended. The instrumental resolution was 0.018 \wn. The estimated absolute calibration uncertainty (1$\sigma$) was 0.003 \wn. The column with $o-c$ displays the deviations between observed values and values calculated by the fitting routine. The branch-label subscripts $e$ and $f$ indicate the upper-state/lower-state symmetry and superscripts $o, p, q, r$ and $s$ denote change in the total angular momentum excluding spin. }
    \end{tablenotes}
  \end{threeparttable}
\end{table}

\subsection{FT-VUV $\boldsymbol{A\leftarrow X}$ system}

\begin{figure*}
  \begin{center}
    \includegraphics[width=\textwidth]{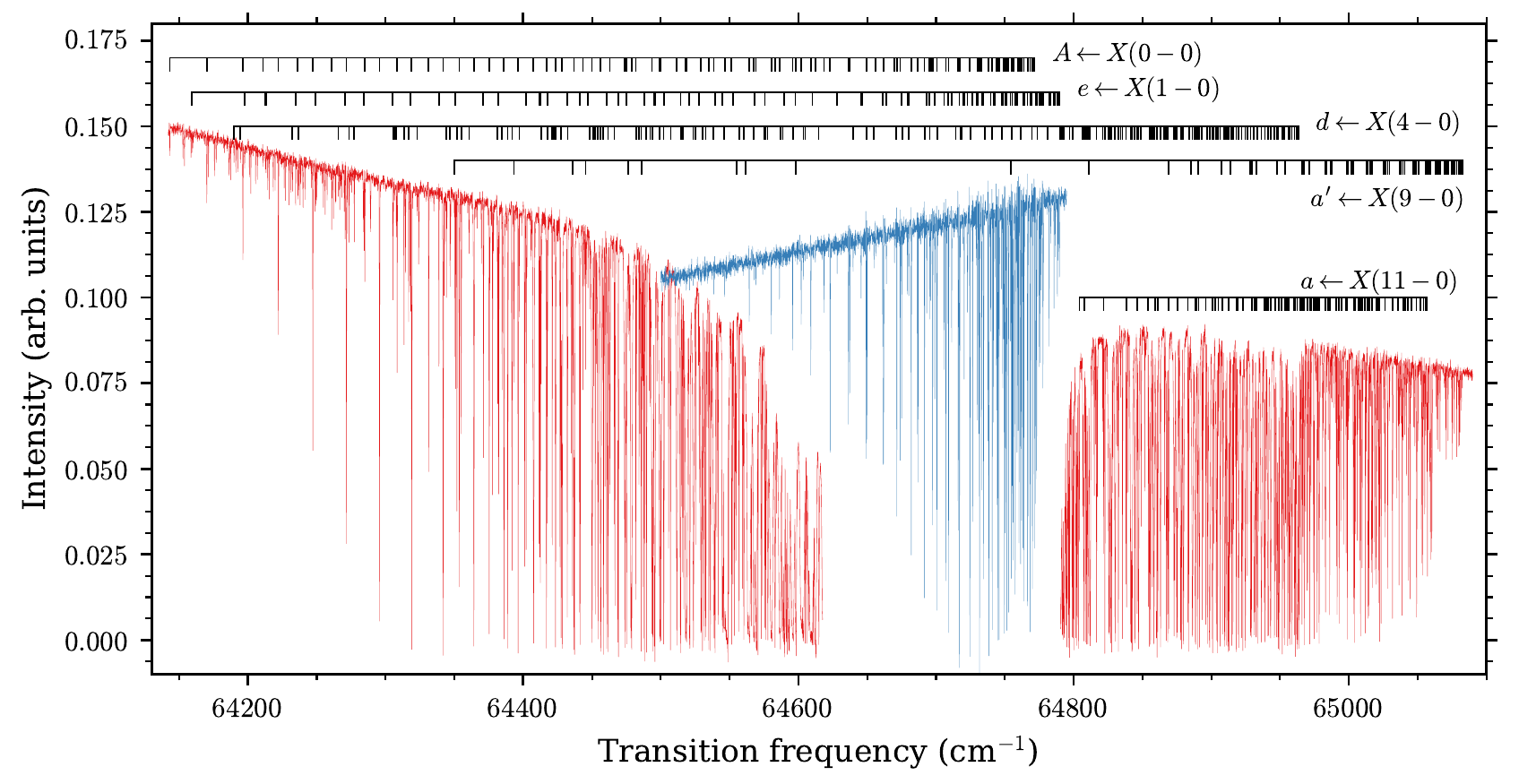}
    \caption{VUV photoabsorption spectra recorded with small and large column densities (blue and red curves, respectively). Observed lines attributed to \CO{} electronic-vibrational bands are indicated.  There is significant absorption due to other CO isotopologues. A variable background intensity slope due to the synchrotron-radiation wavelength dependence is evident.}
    \label{fig:VUV A-X spectrum}
  \end{center}
\end{figure*}

A combination of small and large column-density spectra showing  ${}^{13}$C$^{18}$O \AX$(0,0)$  measured with the FT-VUV is shown in Fig.~\ref{fig:VUV A-X spectrum} (an electronic form of the Fig.~\ref{fig:VUV A-X spectrum}{} source spectrum is provided in the Supplementary Material).
The location of rotational transitions attributed to \AX{}$(0,0)$ are presented in Table~\ref{Table-AX}, while the transition frequencies of four forbidden bands, \eX{}$(1,0)$, \dX{}$(4,0)$, \apX{}$(9,0)$, and \atX{}$(11,0)$, are indicated in this figure and listed in Tables~\ref{Table-dX40} to \ref{Table-apX90}.

\begin{figure}
  \begin{center}
    \includegraphics{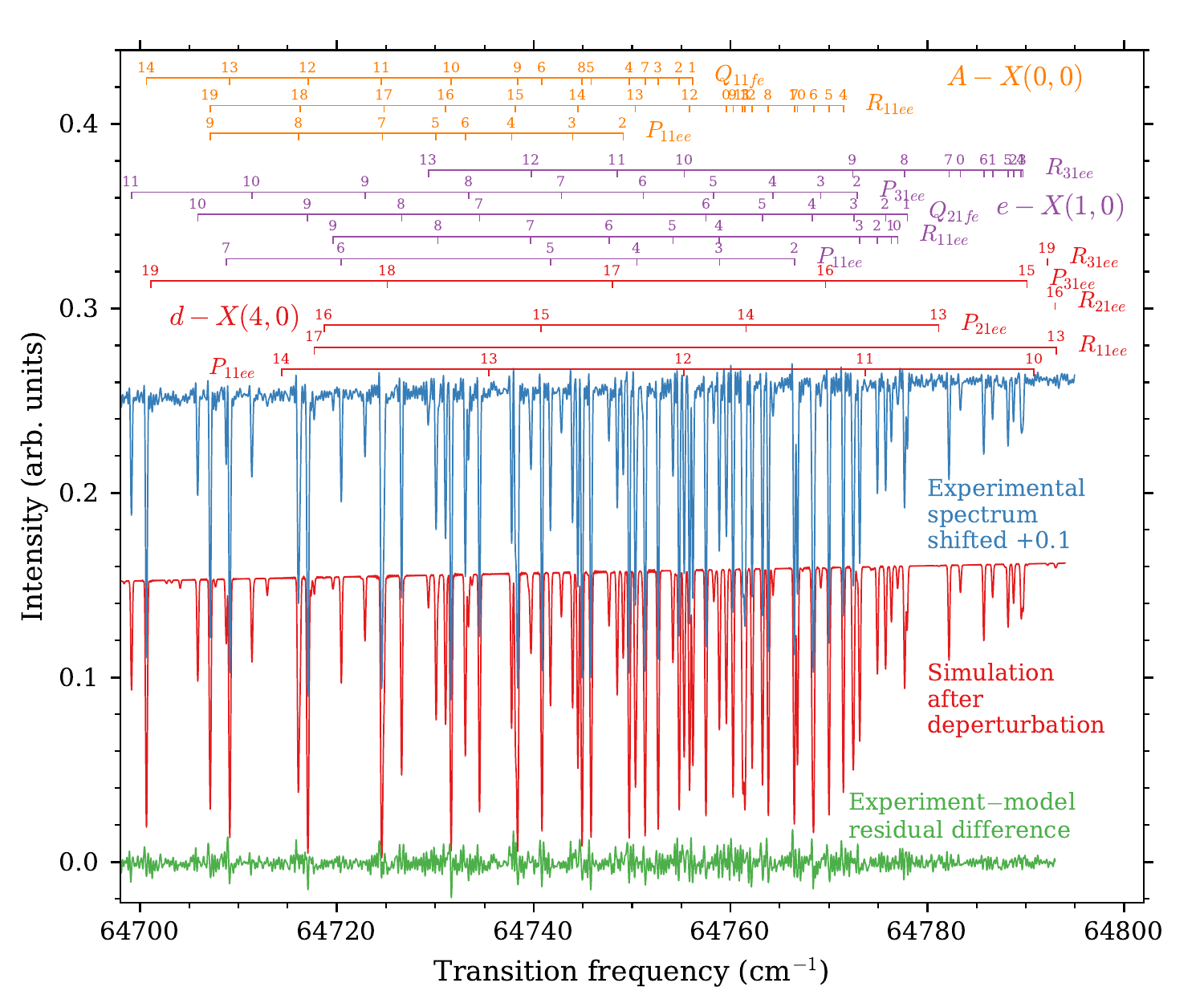}
    \caption {A detailed partial view of the experimental VUV photoabsorption spectra
recorded with large column density, a simulated spectrum employing the
effective Hamiltonian model, and their difference.  The simulation
incorporates several instrumental effects: an assumed column density,
the sloping source intensity, instrumental and Doppler broadening. }
    \label{fig:VUV A-X spectrum detail}
  \end{center}
\end{figure}

A detailed spectrum covering 100\,\wn{} of the high-column-density spectrum is plotted in Fig.~\ref{fig:VUV A-X spectrum detail} and demonstrates the detection of some of these weak transitions.
The assignment of the highly-congested high-density spectrum was greatly facilitated by simultaneously refining the level-interaction model described in Sec.~\ref{sec:Depert}.
Additionally, the purified ${}^{13}$C$^{18}$O sample gas contained minor contamination from the ${}^{13}$C$^{16}$O, ${}^{13}$C$^{17}$O, and ${}^{12}$C$^{18}$O isotopologues, which had to be distinguished from forbidden ${}^{13}$C$^{18}$O transitions. The estimated ratio amounts about $^{13}$C$^{18}$O : $^{13}$C$^{16}$O : $^{13}$C$^{17}$O : $^{12}$C$^{18}$O = 1 : 0.04 : 0.003 : 0.0002. Of the many hundreds of lines observed in the spectra, only some eight lines of non-negligible intensity remain unassigned.

\begin{table*}
\centering
\caption{Measured VUV absorption frequencies\textsuperscript{$a$} of A$\leftarrow$ X$(0,0)$.}
\footnotesize
\begin{threeparttable}
\begin{tabular}{clll}
\label{Table-AX}
\colrule
$J''$              & \multicolumn{1}{c}{$P_{11ee}$}         & \multicolumn{1}{c}{$Q_{11fe}$}         & \multicolumn{1}{c}{$R_{11ee}$}         \\
\colrule
  0                & \multicolumn{1}{c}{--}                 & \multicolumn{1}{c}{--}                 & 64\,759.580(1)     \\
  1                & \multicolumn{1}{c}{--}                 & 64\,756.184(1)     & 64\,761.421(1)     \\
  2                & 64\,749.101(1)     & 64\,754.789(1)     & 64\,762.2024(9)    \\
  3                & 64\,743.958(1)     & 64\,752.652(1)     & 64\,761.511(1)     \\
  4                & 64\,737.7545(9)    & 64\,749.709(1)     & 64\,771.4740(9)    \\
  5                & 64\,730.079(1)     & 64\,745.832(1)     & 64\,770.019(1)     \\
  6                & 64\,733.0599(9)    & 64\,740.845(1)     & 64\,768.460(1)     \\
  7                & 64\,724.623(1)     & 64\,751.331(1)     & 64\,766.489(1)     \\
  8                & 64\,716.085(1)     & 64\,744.926(1)     & 64\,763.855(1)     \\
  9                & 64\,707.136(1)     & 64\,738.376(2)     & 64\,760.268(1)     \\
 10                & 64\,697.527(1)     & 64\,731.619(1)     & 64\,766.806(1)     \\
 11                & 64\,686.966(1)     & 64\,724.529(3)     & 64\,761.245(2)     \\
 12                & 64\,686.533(1)     & 64\,717.08(4)      & 64\,755.838(1)     \\
 13                & 64\,674.003(2)     & 64\,709.122(1)     & 64\,750.333(2)     \\
 14                & 64\,661.631(1)     & 64\,700.653(2)     & 64\,744.493(2)     \\
 15                & 64\,649.165(2)     & 64\,691.629(2)     & 64\,738.160(3)     \\
 16                & 64\,636.367(2)     & 64\,681.990(2)     & 64\,731.048(2)     \\
 17                & 64\,623.078(3)     & 64\,671.514(2)     & 64\,724.791(4)     \\
 18                & 64\,609.016(2)     & 64\,661.872(3)     & 64\,716.247(4)     \\
 19                & 64\,595.813(4)     & 64\,649.880(3)     & 64\,707.115(7)     \\
 20                & 64\,580.327(4)     & 64\,637.314(4)     & 64\,696.222(7)     \\
 21                & 64\,564.258(7)     & 64\,623.004(7)     & 64\,695.29(1)      \\
 22                & 64\,546.433(7)     & 64\,618.606(9)     & 64\,682.16(2)      \\
 23                & 64\,538.57(1)      & 64\,602.018(8)     & 64\,669.84(2)      \\
 24                & 64\,518.52(2)      & 64\,586.24(1)      & 64\,656.21(3)      \\
 25                & 64\,499.28(5)      & 64\,569.13(3)      & \multicolumn{1}{c}{--}                 \\
 26                & 64\,478.74(5)      & \multicolumn{1}{c}{--}                 & \multicolumn{1}{c}{--}                 \\
 27                & 64\,473.75(5)      & \multicolumn{1}{c}{--}                 & \multicolumn{1}{c}{--}                 \\
 28                & 64\,449.96(5)      & 64\,529.24(5)      & 64\,612.50(5)      \\
 29                & 64\,428.20(5)      & 64\,511.49(5)      & 64\,597.86(5)      \\
 30                & 64\,407.46(5)      & 64\,493.50(5)      & 64\,582.88(5)      \\
 31                & 64\,385.95(5)      & 64\,475.05(5)      & 64\,567.42(5)      \\
 32                & 64\,364.10(5)      & 64\,456.36(5)      & 64\,551.43(5)      \\
 33                & 64\,341.77(5)      & 64\,436.86(5)      & 64\,534.866(5)     \\
 34                & 64\,318.93(5)      & 64\,416.86(5)      & 64\,517.679(5)     \\
 35                & 64\,295.519(5)     & 64\,396.28(5)      & 64\,500.120(5)     \\
 36                & 64\,271.495(5)     & 64\,375.27(5)      & 64\,481.810(5)     \\
 37                & 64\,247.107(5)     & 64\,353.537(5)     & 64\,462.935(5)     \\
 38                & 64\,221.977(5)     & 64\,331.243(5)     & 64\,443.461(6)     \\
 39                & 64\,196.290(5)     & 64\,308.360(5)     & 64\,423.39(5)      \\
 40                & 64\,170.014(6)     & 64\,284.886(9)     & \multicolumn{1}{c}{--}                 \\
 41                & 64\,143.15(5)      & 64\,260.81(2)      & \multicolumn{1}{c}{--}                 \\
 42                & \multicolumn{1}{c}{--}                 & 64\,236.15(3)      & \multicolumn{1}{c}{--}                 \\
 43                & \multicolumn{1}{c}{--}                 & 64\,210.88(5)      & \multicolumn{1}{c}{--}                 \\
\colrule
\end{tabular}
\begin{tablenotes}
\begin{footnotesize}
\item[$a$] In units of \wn{} and with $1\sigma$ statistical uncertainties given in parentheses in units of the least-significant digit. The total uncertainty is the sum in quadrature of these and a 0.01\,\wn{} systematic uncertainty.
\end{footnotesize}
\end{tablenotes}
\end{threeparttable}
\end{table*}

Several assumptions were made while analysing these spectra in order to accurately measure the frequencies and absorption depths of blended and weak lines.
The term-value combination differences for $P$- and $R$-branch lines connected to a common upper rotational level were fixed by reference to accurately-known ${}^{13}$C$^{18}$O ground-state term values \cite{Coxon2004} and the ratio of $P(J''-1)/R(J''+1)$ line strengths of all bands was assumed proportional to the corresponding H\"onl-London factors for a ${}^1\Pi-{}^1\Sigma^+$ transition.
This provided an excellent fit to the measured absorption lineshapes and is justified given that the sole source of intensity in this spectral region is the A$\,{}^1\Pi-$X$\,{}^1\Sigma^+$ transition moment, which will maintain its ${}^1\Pi-{}^1\Sigma^+$ character even when redistributed into nominally-forbidden bands.
Additionally, absorption intensities attributed to \apX{}$(9,0)$ were not found to be strongly $J$-dependent once the ground state thermal-population and H\"onl-London factors were factored out.
Then, this band was modelled assuming a quadratic $J(J+1)$ dependence for its band absorption oscillator strengths. Normally-distributed 1$\sigma$-uncertainties for all fitting parameters are estimated from a Hessian matrix computed by the least-squares fitting routine with respect to the fit residual. These uncertainties should be accurate if all assumptions described above are reasonable and the model parameters are not highly correlated. Testing this scheme with an ensemble of synthetic experimental data finds good agreement between estimated uncertainties and statistics of the ensemble.  A minimum uncertainty of 0.05~\wn ~was assumed for frequencies determined from overlapped and saturated lines.
All measured transition frequencies are subject to a common systematic uncertainty associated with the overall frequency calibration. A calibration of the SOLEIL low-pressure spectra was made with respect to the laser-based measurements of Section 3.1. The low-$J$ lines of $^{13}$C$^{18}$O, which are accurately calibrated from the laser spectra (with correction for $\Lambda$-doubling) are all saturated in the high-pressure spectrum and the contaminating lines of ${}^{13}$C${}^{16}$O were instead used for its calibration, with reference to our previous study of this isotopologue \cite{Niu2016}.  The systematic frequency uncertainty is estimated to be 0.01~\wn.

\begin{landscape}
  \begin{table}
    \centering
    \caption{Measured VUV absorption frequencies\textsuperscript{$a$} of d$\leftarrow$ X$(4,0)$.}
    \footnotesize
    \begin{threeparttable}
      \begin{tabular}{clllllllll}
        \label{Table-dX40}
        \colrule
        $J''$              & \multicolumn{1}{c}{$P_{11ee}$}         & \multicolumn{1}{c}{$P_{21ee}$}         & \multicolumn{1}{c}{$P_{31ee}$}         & \multicolumn{1}{c}{$Q_{11fe}$}         & \multicolumn{1}{c}{$Q_{21fe}$}         & \multicolumn{1}{c}{$Q_{31fe}$}         & \multicolumn{1}{c}{$R_{11ee}$}         & \multicolumn{1}{c}{$R_{21ee}$}         & \multicolumn{1}{c}{$R_{31ee}$}         \\
        \colrule
  0                & \multicolumn{1}{c}{--}                 & \multicolumn{1}{c}{--}                 & \multicolumn{1}{c}{--}                 & \multicolumn{1}{c}{--}                 & \multicolumn{1}{c}{--}                 & \multicolumn{1}{c}{--}                 & \multicolumn{1}{c}{--}                 & \multicolumn{1}{c}{--}                 & 64\,962.130(5)     \\
  1                & \multicolumn{1}{c}{--}                 & \multicolumn{1}{c}{--}                 & \multicolumn{1}{c}{--}                 & \multicolumn{1}{c}{--}                 & \multicolumn{1}{c}{--}                 & 64\,958.641(5)     & \multicolumn{1}{c}{--}                 & 64\,926.495(5)     & 64\,963.385(5)     \\
  2                & \multicolumn{1}{c}{--}                 & \multicolumn{1}{c}{--}                 & 64\,951.651(5)     & \multicolumn{1}{c}{--}                 & 64\,919.52(2)      & 64\,956.409(5)     & \multicolumn{1}{c}{--}                 & 64\,926.261(5)     & 64\,963.523(5)     \\
  3                & \multicolumn{1}{c}{--}                 & 64\,909.032(5)     & 64\,945.921(5)     & 64\,879.39(5)      & 64\,915.776(5)     & 64\,953.079(7)     & 64\,887.87(5)      & 64\,924.794(5)     & 64\,962.564(5)     \\
  4                & \multicolumn{1}{c}{--}                 & 64\,901.813(5)     & 64\,939.075(5)     & 64\,873.86(3)      & 64\,910.826(5)     & 64\,948.63(1)      & 64\,884.49(2)      & 64\,922.082(5)     & 64\,960.502(5)     \\
  5                & 64\,856.44(5)      & 64\,893.362(5)     & 64\,931.132(5)     & 64\,866.96(1)      & 64\,904.624(5)     & 64\,943.088(5)     & 64\,879.614(8)     & 64\,918.134(5)     & 64\,957.330(5)     \\
  6                & 64\,846.08(2)      & 64\,883.668(5)     & 64\,922.088(5)     & 64\,858.703(9)     & 64\,897.183(5)     & 64\,936.434(5)     & 64\,873.491(5)     & 64\,912.943(5)     & 64\,953.044(9)     \\
  7                & 64\,834.219(8)     & 64\,872.738(5)     & 64\,911.935(5)     & 64\,849.052(5)     & 64\,888.53(5)      & 64\,928.670(5)     & 64\,866.001(5)     & 64\,906.520(5)     & 64\,947.675(5)     \\
  8                & 64\,821.116(5)     & 64\,860.568(5)     & 64\,900.670(9)     & 64\,838.071(5)     & 64\,878.584(5)     & 64\,919.799(9)     & 64\,857.172(5)     & 64\,898.847(5)     & 64\,941.162(5)     \\
  9                & 64\,806.648(5)     & 64\,847.167(5)     & 64\,888.322(5)     & 64\,825.753(9)     & 64\,867.429(5)     & 64\,909.794(5)     & 64\,847.00(5)      & 64\,889.939(5)     & 64\,933.513(5)     \\
 10                & 64\,790.843(5)     & 64\,832.518(5)     & 64\,874.834(5)     & 64\,812.095(5)     & 64\,855.030(5)     & 64\,898.633(7)     & 64\,835.512(5)     & 64\,879.795(5)     & 64\,924.719(5)     \\
 11                & 64\,773.70(5)      & 64\,816.637(5)     & 64\,860.211(5)     & 64\,797.119(5)     & 64\,841.400(5)     & 64\,886.368(5)     & 64\,822.702(5)     & 64\,868.403(5)     & 64\,914.756(5)     \\
 12                & 64\,755.238(5)     & 64\,799.522(5)     & 64\,844.446(5)     & \multicolumn{1}{c}{--}                 & 64\,826.54(1)      & 64\,872.918(5)     & 64\,808.583(5)     & 64\,855.780(5)     & 64\,903.619(5)     \\
 13                & 64\,735.460(5)     & 64\,781.162(5)     & 64\,827.515(5)     & \multicolumn{1}{c}{--}                 & 64\,810.42(3)      & 64\,858.291(5)     & 64\,793.11(5)      & 64\,841.924(5)     & 64\,891.300(5)     \\
 14                & 64\,714.376(5)     & 64\,761.574(5)     & 64\,809.413(5)     & \multicolumn{1}{c}{--}                 & \multicolumn{1}{c}{--}                 & 64\,842.481(5)     & \multicolumn{1}{c}{--}                 & 64\,826.82(1)      & 64\,877.787(5)     \\
 15                & 64\,691.94(5)      & 64\,740.756(5)     & 64\,790.132(5)     & \multicolumn{1}{c}{--}                 & \multicolumn{1}{c}{--}                 & 64\,825.46(1)      & \multicolumn{1}{c}{--}                 & 64\,810.51(2)      & 64\,863.078(5)     \\
 16                & \multicolumn{1}{c}{--}                 & 64\,718.69(1)      & 64\,769.660(5)     & \multicolumn{1}{c}{--}                 & \multicolumn{1}{c}{--}                 & 64\,807.293(5)     & \multicolumn{1}{c}{--}                 & 64\,792.99(5)      & 64\,847.158(7)     \\
 17                & \multicolumn{1}{c}{--}                 & 64\,695.43(2)      & 64\,747.997(5)     & \multicolumn{1}{c}{--}                 & \multicolumn{1}{c}{--}                 & \multicolumn{1}{c}{--}                 & 64\,717.70(2)      & \multicolumn{1}{c}{--}                 & 64\,830.066(5)     \\
 18                & \multicolumn{1}{c}{--}                 & 64\,670.96(5)      & 64\,725.126(7)     & 64\,654.91(2)      & \multicolumn{1}{c}{--}                 & \multicolumn{1}{c}{--}                 & \multicolumn{1}{c}{--}                 & \multicolumn{1}{c}{--}                 & 64\,811.732(6)     \\
 19                & 64\,588.72(2)      & \multicolumn{1}{c}{--}                 & 64\,701.088(5)     & \multicolumn{1}{c}{--}                 & \multicolumn{1}{c}{--}                 & \multicolumn{1}{c}{--}                 & \multicolumn{1}{c}{--}                 & \multicolumn{1}{c}{--}                 & 64\,792.208(6)     \\
 20                & 64\,560.294(5)     & \multicolumn{1}{c}{--}                 & 64\,675.813(6)     & \multicolumn{1}{c}{--}                 & \multicolumn{1}{c}{--}                 & \multicolumn{1}{c}{--}                 & \multicolumn{1}{c}{--}                 & \multicolumn{1}{c}{--}                 & \multicolumn{1}{c}{--}                 \\
 21                & 64\,530.256(5)     & \multicolumn{1}{c}{--}                 & 64\,649.351(6)     & 64\,575.52(2)      & 64\,639.74(2)      & \multicolumn{1}{c}{--}                 & \multicolumn{1}{c}{--}                 & 64\,680.23(4)      & \multicolumn{1}{c}{--}                 \\
 22                & 64\,498.90(3)      & \multicolumn{1}{c}{--}                 & \multicolumn{1}{c}{--}                 & \multicolumn{1}{c}{--}                 & 64\,603.58(2)      & \multicolumn{1}{c}{--}                 & 64\,595.950(6)     & \multicolumn{1}{c}{--}                 & \multicolumn{1}{c}{--}                 \\
 23                & 64\,466.226(5)     & 64\,523.541(7)     & \multicolumn{1}{c}{--}                 & 64\,515.858(5)     & 64\,577.667(7)     & \multicolumn{1}{c}{--}                 & 64\,567.667(9)     & \multicolumn{1}{c}{--}                 & \multicolumn{1}{c}{--}                 \\
 24                & 64\,432.312(6)     & 64\,494.124(5)     & \multicolumn{1}{c}{--}                 & 64\,484.118(5)     & \multicolumn{1}{c}{--}                 & \multicolumn{1}{c}{--}                 & 64\,538.11(2)      & 64\,604.955(5)     & \multicolumn{1}{c}{--}                 \\
 25                & 64\,397.114(9)     & 64\,461.670(5)     & 64\,532.644(5)     & 64\,451.103(7)     & \multicolumn{1}{c}{--}                 & \multicolumn{1}{c}{--}                 & 64\,507.29(2)      & 64\,576.257(5)     & \multicolumn{1}{c}{--}                 \\
 26                & 64\,360.65(2)      & 64\,427.491(5)     & 64\,502.030(5)     & \multicolumn{1}{c}{--}                 & 64\,485.796(5)     & \multicolumn{1}{c}{--}                 & \multicolumn{1}{c}{--}                 & 64\,546.224(5)     & 64\,614.732(5)     \\
 27                & 64\,322.92(2)      & 64\,391.889(5)     & 64\,454.221(5)     & 64\,381.27(3)      & 64\,452.312(5)     & \multicolumn{1}{c}{--}                 & \multicolumn{1}{c}{--}                 & 64\,514.887(5)     & 64\,586.878(5)     \\
 28                & \multicolumn{1}{c}{--}                 & 64\,354.959(5)     & 64\,423.467(5)     & 64\,344.48(3)      & 64\,417.534(5)     & 64\,489.539(5)     & \multicolumn{1}{c}{--}                 & 64\,482.271(6)     & 64\,556.877(5)     \\
 29                & \multicolumn{1}{c}{--}                 & 64\,316.732(5)     & 64\,388.723(5)     & 64\,306.41(5)      & 64\,381.463(6)     & 64\,456.08(1)      & \multicolumn{1}{c}{--}                 & 64\,448.38(1)      & 64\,525.287(5)     \\
 30                & \multicolumn{1}{c}{--}                 & 64\,277.231(6)     & 64\,351.837(5)     & \multicolumn{1}{c}{--}                 & 64\,344.138(9)     & 64\,421.047(5)     & \multicolumn{1}{c}{--}                 & 64\,413.23(3)      & 64\,492.275(6)     \\
 31                & \multicolumn{1}{c}{--}                 & 64\,236.46(1)      & 64\,313.371(5)     & \multicolumn{1}{c}{--}                 & 64\,305.56(2)      & 64\,384.606(5)     & \multicolumn{1}{c}{--}                 & \multicolumn{1}{c}{--}                 & 64\,457.94(2)      \\
 32                & \multicolumn{1}{c}{--}                 & 64\,194.44(3)      & 64\,273.489(6)     & \multicolumn{1}{c}{--}                 & 64\,265.71(3)      & 64\,346.84(1)      & \multicolumn{1}{c}{--}                 & \multicolumn{1}{c}{--}                 & 64\,422.31(2)      \\
 33                & \multicolumn{1}{c}{--}                 & \multicolumn{1}{c}{--}                 & 64\,232.29(2)      & \multicolumn{1}{c}{--}                 & \multicolumn{1}{c}{--}                 & 64\,307.78(2)      & \multicolumn{1}{c}{--}                 & \multicolumn{1}{c}{--}                 & \multicolumn{1}{c}{--}                 \\
 34                & \multicolumn{1}{c}{--}                 & \multicolumn{1}{c}{--}                 & 64\,189.81(2)      & \multicolumn{1}{c}{--}                 & \multicolumn{1}{c}{--}                 & \multicolumn{1}{c}{--}                 & \multicolumn{1}{c}{--}                 & \multicolumn{1}{c}{--}                 & \multicolumn{1}{c}{--}                 \\
        \colrule
      \end{tabular}
      \begin{tablenotes}
        \begin{footnotesize}
          \item[$a$] In units of \wn{} and with $1\sigma$ statistical uncertainties given in parentheses in units of the least-significant digit. The total uncertainty is the sum in quadrature of these and a 0.01\,\wn{} systematic uncertainty.
        \end{footnotesize}
      \end{tablenotes}
    \end{threeparttable}
  \end{table}
\end{landscape}

\begin{table*}
\centering
\caption{Measured VUV absorption frequencies\textsuperscript{$a$} of e$\leftarrow$ X$(1,0)$.}
\footnotesize
\begin{threeparttable}
\begin{tabular}{clllll}
\label{Table-eX10}
\colrule
$J''$              & \multicolumn{1}{c}{$P_{11ee}$}         & \multicolumn{1}{c}{$P_{31ee}$}         & \multicolumn{1}{c}{$Q_{21fe}$}         & \multicolumn{1}{c}{$R_{11ee}$}         & \multicolumn{1}{c}{$R_{31ee}$}         \\
\colrule
  0                & \multicolumn{1}{c}{--}                 & \multicolumn{1}{c}{--}                 & \multicolumn{1}{c}{--}                 & 64\,776.99(2)      & 64\,783.37(1)      \\
  1                & \multicolumn{1}{c}{--}                 & \multicolumn{1}{c}{--}                 & 64\,777.968(7)     & 64\,776.347(7)     & 64\,786.637(9)     \\
  2                & 64\,766.51(2)      & 64\,772.89(1)      & 64\,775.773(4)     & 64\,774.932(3)     & 64\,788.766(8)     \\
  3                & 64\,758.884(7)     & 64\,769.173(9)     & 64\,772.529(5)     & 64\,773.138(2)     & 64\,789.725(8)     \\
  4                & 64\,750.484(3)     & 64\,764.318(8)     & 64\,768.333(4)     & 64\,758.864(4)     & 64\,789.548(9)     \\
  5                & 64\,741.707(2)     & 64\,758.293(8)     & 64\,763.261(2)     & 64\,754.149(4)     & 64\,788.210(5)     \\
  6                & 64\,720.450(4)     & 64\,751.134(9)     & 64\,757.511(2)     & 64\,747.650(7)     & 64\,785.746(4)     \\
  7                & 64\,708.754(4)     & 64\,742.814(5)     & 64\,734.499(2)     & 64\,739.66(2)      & 64\,782.209(3)     \\
  8                & 64\,695.275(7)     & 64\,733.372(4)     & 64\,726.583(2)     & 64\,730.26(1)      & 64\,777.693(3)     \\
  9                & 64\,680.31(2)      & 64\,722.856(3)     & 64\,717.0(1)       & 64\,719.62(3)      & 64\,772.427(3)     \\
 10                & 64\,663.94(1)      & 64\,711.364(3)     & 64\,705.864(4)     & \multicolumn{1}{c}{--}                 & 64\,755.303(2)     \\
 11                & 64\,646.32(3)      & 64\,699.125(3)     & 64\,693.238(7)     & \multicolumn{1}{c}{--}                 & 64\,748.495(3)     \\
 12                & \multicolumn{1}{c}{--}                 & 64\,675.030(2)     & 64\,679.210(9)     & \multicolumn{1}{c}{--}                 & 64\,739.748(8)     \\
 13                & \multicolumn{1}{c}{--}                 & 64\,661.254(3)     & 64\,663.83(7)      & \multicolumn{1}{c}{--}                 & 64\,729.304(9)     \\
 14                & \multicolumn{1}{c}{--}                 & 64\,645.541(8)     & \multicolumn{1}{c}{--}                 & \multicolumn{1}{c}{--}                 & \multicolumn{1}{c}{--}                 \\
 15                & \multicolumn{1}{c}{--}                 & 64\,628.136(9)     & \multicolumn{1}{c}{--}                 & \multicolumn{1}{c}{--}                 & \multicolumn{1}{c}{--}                 \\
 16                & 64\,539.422(5)     & \multicolumn{1}{c}{--}                 & \multicolumn{1}{c}{--}                 & 64\,610.166(5)     & \multicolumn{1}{c}{--}                 \\
 17                & 64\,514.39(1)      & \multicolumn{1}{c}{--}                 & \multicolumn{1}{c}{--}                 & 64\,589.644(5)     & \multicolumn{1}{c}{--}                 \\
 18                & 64\,488.135(5)     & \multicolumn{1}{c}{--}                 & \multicolumn{1}{c}{--}                 & 64\,567.926(5)     & \multicolumn{1}{c}{--}                 \\
 19                & 64\,460.666(5)     & 64\,544.628(5)     & \multicolumn{1}{c}{--}                 & 64\,544.987(5)     & \multicolumn{1}{c}{--}                 \\
 20                & 64\,432.006(5)     & 64\,520.54(3)      & \multicolumn{1}{c}{--}                 & 64\,520.834(5)     & \multicolumn{1}{c}{--}                 \\
 21                & 64\,402.130(5)     & 64\,495.26(5)      & 64\,495.717(5)     & 64\,495.465(5)     & 64\,597.716(5)     \\
 22                & 64\,371.045(5)     & 64\,468.744(5)     & 64\,469.181(5)     & 64\,468.888(7)     & 64\,575.677(5)     \\
 23                & 64\,338.749(5)     & 64\,441.000(5)     & 64\,441.423(5)     & 64\,441.09(1)      & 64\,552.405(5)     \\
 24                & 64\,305.250(7)     & 64\,412.039(5)     & 64\,412.447(5)     & 64\,412.08(2)      & 64\,527.912(5)     \\
 25                & 64\,270.54(1)      & 64\,381.851(5)     & 64\,382.257(5)     & 64\,381.87(2)      & 64\,502.191(8)     \\
 26                & 64\,234.61(2)      & 64\,350.449(5)     & 64\,350.848(5)     & 64\,350.47(3)      & 64\,475.11(2)      \\
 27                & 64\,197.50(2)      & 64\,317.824(8)     & 64\,318.234(8)     & \multicolumn{1}{c}{--}                 & 64\,447.10(1)      \\
 28                & 64\,159.20(3)      & 64\,283.84(2)      & 64\,284.40(1)      & \multicolumn{1}{c}{--}                 & \multicolumn{1}{c}{--}                 \\
 29                & \multicolumn{1}{c}{--}                 & 64\,248.95(1)      & 64\,249.34(2)      & \multicolumn{1}{c}{--}                 & \multicolumn{1}{c}{--}                 \\
 30                & \multicolumn{1}{c}{--}                 & 64\,212.66(5)      & 64\,213.09(4)      & \multicolumn{1}{c}{--}                 & \multicolumn{1}{c}{--}                 \\
\colrule
\end{tabular}
\begin{tablenotes}
\begin{footnotesize}
\item[$a$] In units of \wn{} and with $1\sigma$ statistical uncertainties given in parentheses in units of the least-significant digit. The total uncertainty is the sum in quadrature of these and a 0.01\,\wn{} systematic uncertainty.
\end{footnotesize}
\end{tablenotes}
\end{threeparttable}
\end{table*}

\begin{landscape}
\begin{table}
\centering
\caption{Measured VUV absorption frequencies\textsuperscript{$a$} of a$\leftarrow$ X$(11,0)$.}
\footnotesize
\begin{threeparttable}
\begin{tabular}{clllllllll}
\label{Table-aX110}
\colrule
$J''$              & \multicolumn{1}{c}{$P_{11ee}$}         & \multicolumn{1}{c}{$P_{21ee}$}         & \multicolumn{1}{c}{$P_{31ee}$}         & \multicolumn{1}{c}{$Q_{11fe}$}         & \multicolumn{1}{c}{$Q_{21fe}$}         & \multicolumn{1}{c}{$Q_{31fe}$}         & \multicolumn{1}{c}{$R_{11ee}$}         & \multicolumn{1}{c}{$R_{21ee}$}         & \multicolumn{1}{c}{$R_{31ee}$}         \\
\colrule
  0                & \multicolumn{1}{c}{--}                 & \multicolumn{1}{c}{--}                 & \multicolumn{1}{c}{--}                 & \multicolumn{1}{c}{--}                 & \multicolumn{1}{c}{--}                 & \multicolumn{1}{c}{--}                 & 64\,976.03(2)      & 65\,019.973(5)     & \multicolumn{1}{c}{--}                 \\
  1                & \multicolumn{1}{c}{--}                 & \multicolumn{1}{c}{--}                 & \multicolumn{1}{c}{--}                 & \multicolumn{1}{c}{--}                 & 65\,016.477(5)     & \multicolumn{1}{c}{--}                 & 64\,977.677(5)     & 65\,021.675(5)     & \multicolumn{1}{c}{--}                 \\
  2                & 64\,965.55(2)      & 65\,009.495(5)     & \multicolumn{1}{c}{--}                 & 64\,972.523(5)     & 65\,014.692(5)     & \multicolumn{1}{c}{--}                 & 64\,978.389(5)     & 65\,022.467(5)     & \multicolumn{1}{c}{--}                 \\
  3                & 64\,960.214(5)     & 65\,004.211(5)     & \multicolumn{1}{c}{--}                 & 64\,969.713(5)     & 65\,012.030(5)     & \multicolumn{1}{c}{--}                 & 64\,978.149(5)     & 65\,022.403(5)     & \multicolumn{1}{c}{--}                 \\
  4                & 64\,953.942(5)     & 64\,998.019(5)     & \multicolumn{1}{c}{--}                 & 64\,965.952(5)     & 65\,008.503(5)     & 65\,042.68(3)      & 64\,976.955(5)     & 65\,021.490(5)     & \multicolumn{1}{c}{--}                 \\
  5                & 64\,946.717(5)     & 64\,990.971(5)     & \multicolumn{1}{c}{--}                 & 64\,961.220(5)     & 65\,004.123(5)     & 65\,039.63(1)      & 64\,974.788(5)     & 65\,019.741(5)     & 65\,056.84(2)      \\
  6                & 64\,938.541(5)     & 64\,983.076(5)     & \multicolumn{1}{c}{--}                 & 64\,955.517(5)     & 64\,998.926(5)     & 65\,035.88(1)      & 64\,971.650(5)     & 65\,017.176(5)     & 65\,055.95(2)      \\
  7                & 64\,929.393(5)     & 64\,974.346(5)     & 65\,011.45(2)      & 64\,948.86(5)      & 64\,992.905(5)     & 65\,031.54(5)      & 64\,967.527(5)     & 65\,013.805(5)     & 65\,054.28(1)      \\
  8                & 64\,919.275(5)     & 64\,964.801(5)     & 65\,003.57(2)      & 64\,941.16(5)      & 64\,986.083(5)     & 65\,026.48(2)      & 64\,962.443(5)     & 65\,009.636(5)     & 65\,052.19(4)      \\
  9                & 64\,908.175(5)     & 64\,954.452(5)     & 64\,994.92(1)      & 64\,932.56(3)      & 64\,978.49(1)      & 65\,020.66(1)      & 64\,956.407(5)     & 65\,004.664(5)     & 65\,049.09(3)      \\
 10                & 64\,896.114(5)     & 64\,943.307(5)     & 64\,985.86(4)      & 64\,922.981(5)     & 64\,970.063(5)     & \multicolumn{1}{c}{--}                 & 64\,949.416(5)     & 64\,998.920(6)     & 65\,045.36(1)      \\
 11                & 64\,883.105(5)     & 64\,931.362(5)     & 64\,975.79(3)      & 64\,912.48(1)      & 64\,960.872(5)     & 65\,006.97(1)      & 64\,941.504(5)     & 64\,992.371(5)     & 65\,040.92(2)      \\
 12                & 64\,869.144(5)     & 64\,918.647(6)     & 64\,965.09(1)      & 64\,901.008(9)     & 64\,950.926(6)     & \multicolumn{1}{c}{--}                 & 64\,932.678(5)     & 64\,985.042(6)     & 65\,035.71(5)      \\
 13                & 64\,854.263(5)     & 64\,905.130(5)     & 64\,953.68(2)      & 64\,888.65(5)      & 64\,940.138(5)     & \multicolumn{1}{c}{--}                 & 64\,922.942(5)     & 64\,976.887(9)     & \multicolumn{1}{c}{--}                 \\
 14                & 64\,838.471(5)     & 64\,890.836(6)     & 64\,941.50(5)      & 64\,875.386(7)     & \multicolumn{1}{c}{--}                 & \multicolumn{1}{c}{--}                 & 64\,912.316(5)     & 64\,967.73(2)      & \multicolumn{1}{c}{--}                 \\
 15                & 64\,821.774(5)     & 64\,875.719(9)     & \multicolumn{1}{c}{--}                 & 64\,861.204(7)     & \multicolumn{1}{c}{--}                 & \multicolumn{1}{c}{--}                 & \multicolumn{1}{c}{--}                 & \multicolumn{1}{c}{--}                 & \multicolumn{1}{c}{--}                 \\
 16                & 64\,804.190(5)     & 64\,859.60(2)      & \multicolumn{1}{c}{--}                 & 64\,846.127(5)     & 64\,903.10(2)      & \multicolumn{1}{c}{--}                 & \multicolumn{1}{c}{--}                 & \multicolumn{1}{c}{--}                 & \multicolumn{1}{c}{--}                 \\
 17                & \multicolumn{1}{c}{--}                 & \multicolumn{1}{c}{--}                 & \multicolumn{1}{c}{--}                 & \multicolumn{1}{c}{--}                 & 64\,889.21(2)      & \multicolumn{1}{c}{--}                 & \multicolumn{1}{c}{--}                 & \multicolumn{1}{c}{--}                 & \multicolumn{1}{c}{--}                 \\
 18                & \multicolumn{1}{c}{--}                 & \multicolumn{1}{c}{--}                 & \multicolumn{1}{c}{--}                 & \multicolumn{1}{c}{--}                 &\multicolumn{1}{c}{--}      & \multicolumn{1}{c}{--}                 & \multicolumn{1}{c}{--}                 & \multicolumn{1}{c}{--}                 & \multicolumn{1}{c}{--}                 \\
 19                & \multicolumn{1}{c}{--}                 & \multicolumn{1}{c}{--}                 & \multicolumn{1}{c}{--}                 & \multicolumn{1}{c}{--}                 & 64\,858.97(4)      & \multicolumn{1}{c}{--}                 & \multicolumn{1}{c}{--}                 & \multicolumn{1}{c}{--}                 & \multicolumn{1}{c}{--}                 \\
\colrule
\end{tabular}
\begin{tablenotes}
\begin{footnotesize}
\item[$a$] In units of \wn{} and with $1\sigma$ statistical uncertainties given in parentheses in units of the least-significant digit. The total uncertainty is the sum in quadrature of these and a 0.01\,\wn{} systematic uncertainty.
\end{footnotesize}
\end{tablenotes}
\end{threeparttable}
\end{table}
\end{landscape}

\begin{table*}
\centering
\caption{Measured VUV absorption frequencies\textsuperscript{$a$} of a$'\leftarrow$ X$(9,0)$.}
\footnotesize
\begin{threeparttable}
\begin{tabular}{cllll}
\label{Table-apX90}
\colrule
$J''$              & \multicolumn{1}{c}{$P_{21ee}$}         & \multicolumn{1}{c}{$Q_{11fe}$}         & \multicolumn{1}{c}{$Q_{31fe}$}         & \multicolumn{1}{c}{$R_{21ee}$}         \\
\colrule
  0                & \multicolumn{1}{c}{--}                 & \multicolumn{1}{c}{--}                 & \multicolumn{1}{c}{--}                 & 65\,079.97(1)      \\
  1                & \multicolumn{1}{c}{--}                 & 65\,074.91(1)      & 65\,082.50(2)      & 65\,080.794(8)     \\
  2                & 65\,069.49(1)      & 65\,070.29(1)      & 65\,081.82(1)      & 65\,080.288(7)     \\
  3                & 65\,063.331(8)     & 65\,064.224(8)     & 65\,079.88(1)      & 65\,078.447(6)     \\
  4                & 65\,055.840(7)     & 65\,056.768(9)     & 65\,076.660(9)     & 65\,075.276(5)     \\
  5                & 65\,047.015(6)     & 65\,047.989(7)     & 65\,072.105(9)     & 65\,070.777(5)     \\
  6                & 65\,036.862(5)     & 65\,037.897(7)     & 65\,066.253(8)     & 65\,064.945(6)     \\
  7                & 65\,025.381(5)     & 65\,026.36(3)      & 65\,059.066(8)     & 65\,057.837(6)     \\
  8                & 65\,012.570(6)     & 65\,013.54(1)      & 65\,050.552(8)     & 65\,049.10(1)      \\
  9                & 64\,998.484(6)     & 64\,999.40(1)      & 65\,040.710(9)     & 65\,039.357(9)     \\
 10                & 64\,982.77(1)      & 64\,983.93(1)      & 65\,029.550(8)     & 65\,028.19(2)      \\
 11                & 64\,966.055(9)     & 64\,967.15(2)      & 65\,017.06(1)      & 65\,015.66(1)      \\
 12                & 64\,947.92(2)      & \multicolumn{1}{c}{--}                 & 65\,003.28(2)      & 65\,001.88(1)      \\
 13                & 64\,928.41(1)      & \multicolumn{1}{c}{--}                 & 64\,987.80(1)      & 64\,986.71(2)      \\
 14                & 64\,907.68(1)      & \multicolumn{1}{c}{--}                 & 64\,971.44(2)      & \multicolumn{1}{c}{--}                 \\
 15                & 64\,885.55(2)      & \multicolumn{1}{c}{--}                 & 64\,953.60(2)      & \multicolumn{1}{c}{--}                 \\
 16                & \multicolumn{1}{c}{--}                 & \multicolumn{1}{c}{--}                 & \multicolumn{1}{c}{--}                 & 64\,933.03(5)      \\
 17                & \multicolumn{1}{c}{--}                 & \multicolumn{1}{c}{--}                 & 64\,913.88(2)      & \multicolumn{1}{c}{--}                 \\
 18                & 64\,811.00(5)      & \multicolumn{1}{c}{--}                 & \multicolumn{1}{c}{--}                 & \multicolumn{1}{c}{--}                 \\
 19                & \multicolumn{1}{c}{--}                 & \multicolumn{1}{c}{--}                 & 64\,868.90(5)      & \multicolumn{1}{c}{--}                 \\
 26                & \multicolumn{1}{c}{--}                 & 64\,554.926(7)     & \multicolumn{1}{c}{--}                 & \multicolumn{1}{c}{--}                 \\
 28                & 64\,476.41(1)      & \multicolumn{1}{c}{--}                 & \multicolumn{1}{c}{--}                 & 64\,598.320(5)     \\
 29                & 64\,435.837(5)     & \multicolumn{1}{c}{--}                 & \multicolumn{1}{c}{--}                 & 64\,561.84(2)      \\
 30                & 64\,393.280(5)     & \multicolumn{1}{c}{--}                 & \multicolumn{1}{c}{--}                 & \multicolumn{1}{c}{--}                 \\
 31                & 64\,349.92(2)      & \multicolumn{1}{c}{--}                 & 64\,486.221(7)     & \multicolumn{1}{c}{--}                 \\
 32                & \multicolumn{1}{c}{--}                 & \multicolumn{1}{c}{--}                 & 64\,445.434(7)     & \multicolumn{1}{c}{--}                 \\
\colrule
\end{tabular}
\begin{tablenotes}
\begin{footnotesize}
\item[$a$] In units of \wn{} and with $1\sigma$ statistical uncertainties given in parentheses in units of the least-significant digit. The total uncertainty is the sum in quadrature of these and a 0.01\,\wn{} systematic uncertainty.
\end{footnotesize}
\end{tablenotes}
\end{threeparttable}
\end{table*}

To conclude we find no need to assign lines to additional electronic transitions, beyond the ones known to exist in this energy range of CO. All lines that were postulated to belong to the $P$ and $R$-transitions in a 1$^1\Sigma^+$ - X$^1\Sigma^+$ band system and to $Q$-branch transitions in a 2$^1\Pi$ - X$^1\Sigma^+$ band system in a previous study~\cite{Lemaire2016} could be assigned to \AX$(0, 0)$ lines and perturber lines of \CO{}  as found in the present paper.

\subsection{FT-VUV $\boldsymbol{B\leftarrow X}$ system}

A spectrum of \BX$(0,0)$ was recorded with the FT-VUV setup at SOLEIL while \CO{} flowed through a windowless cell heated to approximately 1000\,K \cite{Niu2015a}.
The column density for this measurement was approximately \np[cm^{-2}]{7e15} and overlapping \BX$(0,0)$ absorption from the ${}^{12}$C$^{16}$O, ${}^{13}$C$^{16}$O, ${}^{13}$C$^{17}$O, and ${}^{12}$C$^{18}$O isotopologues, as well as the \BX$(1,0)$ band of \CO{}, had to be included in the analysis of this spectrum, with observed rotational levels up to $J=51$.

\begin{table}
\centering
\caption{Measured FT-VUV absorption frequencies\textsuperscript{$a$} of B$\leftarrow$ X$(0,0)$.}
  \label{Table-BX}
\footnotesize
\begin{threeparttable}
\begin{tabular}{cll}
\colrule
$J''$              & \multicolumn{1}{c}{$P_{11ee}$}         & \multicolumn{1}{c}{$R_{11ee}$}         \\
\colrule
  0                & \multicolumn{1}{c}{--}                 & 86\,920.896(1)     \\
  1                & 86\,913.862(2)     & 86\,924.483(1)     \\
  2                & 86\,910.418(1)     & 86\,928.1164(9)    \\
  3                & 86\,907.020(1)     & 86\,931.7944(8)    \\
  4                & 86\,903.6685(9)    & 86\,935.5212(8)    \\
  5                & 86\,900.3628(8)    & 86\,939.2919(7)    \\
  6                & 86\,897.1071(8)    & 86\,943.1082(7)    \\
  7                & 86\,893.8966(7)    & 86\,946.9702(7)    \\
  8                & 86\,890.7332(7)    & 86\,950.8773(7)    \\
  9                & 86\,887.6174(7)    & 86\,954.8299(7)    \\
 10                & 86\,884.5487(7)    & 86\,958.8257(8)    \\
 11                & 86\,881.5279(7)    & 86\,962.8701(8)    \\
 12                & 86\,878.5527(8)    & 86\,966.9544(8)    \\
 13                & 86\,875.6290(8)    & 86\,971.0857(8)    \\
 14                & 86\,872.7481(8)    & 86\,975.2595(9)    \\
 15                & 86\,869.9176(8)    & 86\,979.4751(9)    \\
 16                & 86\,867.1331(9)    & 86\,983.694(1)     \\
 17                & 86\,864.3941(9)    & 86\,988.035(1)     \\
 18                & 86\,861.662(1)     & 86\,992.390(1)     \\
 19                & 86\,859.057(1)     & 86\,996.776(1)     \\
 20                & 86\,856.470(1)     & 87\,001.205(1)     \\
 21                & 86\,853.919(1)     & 87\,005.677(1)     \\
 22                & 86\,851.416(1)     & 87\,010.189(1)     \\
 23                & 86\,848.961(1)     & 87\,014.744(2)     \\
 24                & 86\,846.551(1)     & 87\,019.336(2)     \\
 25                & 86\,844.190(2)     & 87\,023.973(2)     \\
 26                & 86\,841.872(2)     & 87\,028.651(2)     \\
 27                & 86\,839.606(2)     & 87\,033.359(2)     \\
 28                & 86\,837.386(2)     & 87\,038.106(2)     \\
 29                & 86\,835.204(2)     & 87\,042.899(2)     \\
 30                & 86\,833.067(2)     & 87\,047.714(2)     \\
 31                & 86\,830.982(2)     & 87\,052.576(2)     \\
 32                & 86\,828.928(2)     & 87\,057.485(3)     \\
 33                & 86\,826.929(2)     & 87\,062.425(3)     \\
 34                & 86\,824.984(3)     & 87\,067.412(3)     \\
 35                & 86\,823.078(3)     & 87\,072.399(4)     \\
 36                & 86\,821.227(3)     & 87\,077.436(4)     \\
 37                & 86\,819.385(4)     & 87\,082.517(5)     \\
 38                & 86\,817.602(4)     & 87\,087.625(5)     \\
 39                & 86\,815.872(5)     & 87\,092.760(6)     \\
 40                & 86\,814.178(5)     & 87\,097.933(7)     \\
 41                & 86\,812.520(6)     & 87\,103.116(9)     \\
 42                & 86\,810.911(7)     & 87\,108.36(1)      \\
 43                & 86\,809.321(9)     & 87\,113.63(1)      \\
 44                & 86\,807.80(1)      & 87\,118.92(1)      \\
 45                & 86\,806.32(1)      & 87\,124.27(2)      \\
 46                & 86\,804.87(1)      & 87\,129.58(2)      \\
 47                & 86\,803.49(2)      & 87\,134.94(3)      \\
 48                & 86\,802.08(2)      & 87\,140.37(3)      \\
 49                & 86\,800.73(3)      & 87\,145.74(4)      \\
 50                & 86\,799.47(3)      & \multicolumn{1}{c}{--}                 \\
 51                & 86\,798.15(4)      & \multicolumn{1}{c}{--}                 \\
\colrule
\end{tabular}
\begin{tablenotes}
\begin{footnotesize}
\item[$a$] In units of \wn{} and with $1\sigma$ statistical uncertainties given in parentheses in units of the least-significant digit. The total uncertainty is the sum in quadrature of these and a 0.01\,\wn{} systematic uncertainty.
\end{footnotesize}
\end{tablenotes}
\end{threeparttable}
\end{table}

\begin{figure}
\begin{center}
  \includegraphics{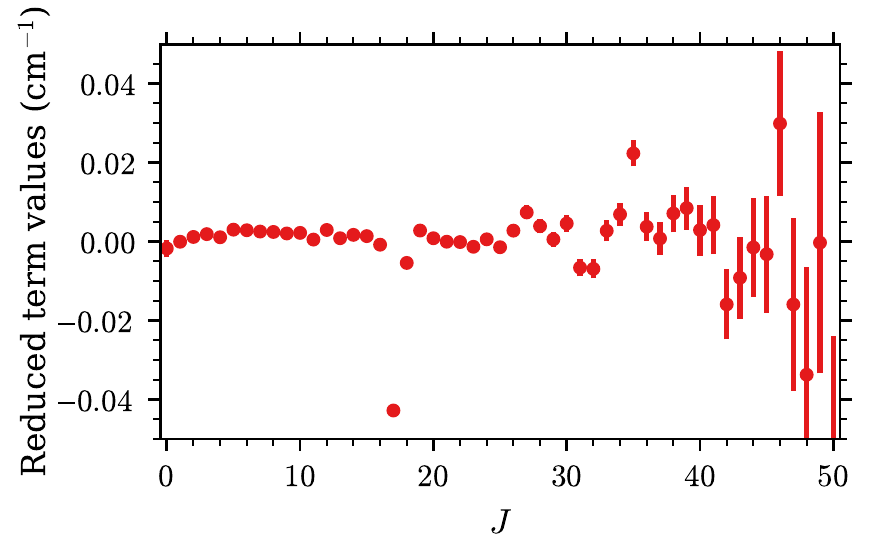}
  \caption{Experimental term values of the \CO{} B$(v=0)$ level after subtraction by a best-fitting second-order polynomial in terms of $J(J+1)$. The error bars indicate $1\sigma$ statistical fitting uncertainties and term-value deviations at some values of $J$ exceeding these uncertainties are indicative of perturbations.}
  \label{fig:B(0) reduced T}
\end{center}
\end{figure}

Measured \BX$(0,0)$ transition frequencies and term values are listed in Tables~\ref{Table-BX} and \ref{B-term}, respectively. The purpose of measuring this band is twofold.
First, transition frequencies from the  \BA$(0,0)$ FT-VIS spectrum can be converted to absolute A($v=0$)-state term values when combined with the \BX$(0,0)$ data and known ground state levels. Second, there occur small perturbations in the \BX$(0,0)$ spectrum leading to the B$(v=0)$ term energy shifts shown in Fig.~\ref{fig:B(0) reduced T} at $J=17$, and between 30 and 35. Care was taken that these were not misinterpreted as shifts of \As($v=0$) levels while analysing the \BA$(0,0)$ spectrum. The frequency calibration of this spectrum was made with respect to atomic lines contaminating the spectrum and adopted NIST frequencies for Xe (lines at 83\,889.97, 85\,440.02, and 90\,032.18\,cm$^{-1}$), Kr (85\,846.71\,cm$^{-1}$), and O (86\,794.15\,cm$^{-1}$). This calibration is estimated to impart a systematic uncertainty of 0.01 \,cm$^{-1}$ to the measured line frequencies of this band.

\section{Deperturbation analysis}
\label{sec:Depert}

\begin{figure}
  \begin{center}
    \includegraphics[width=\textwidth]{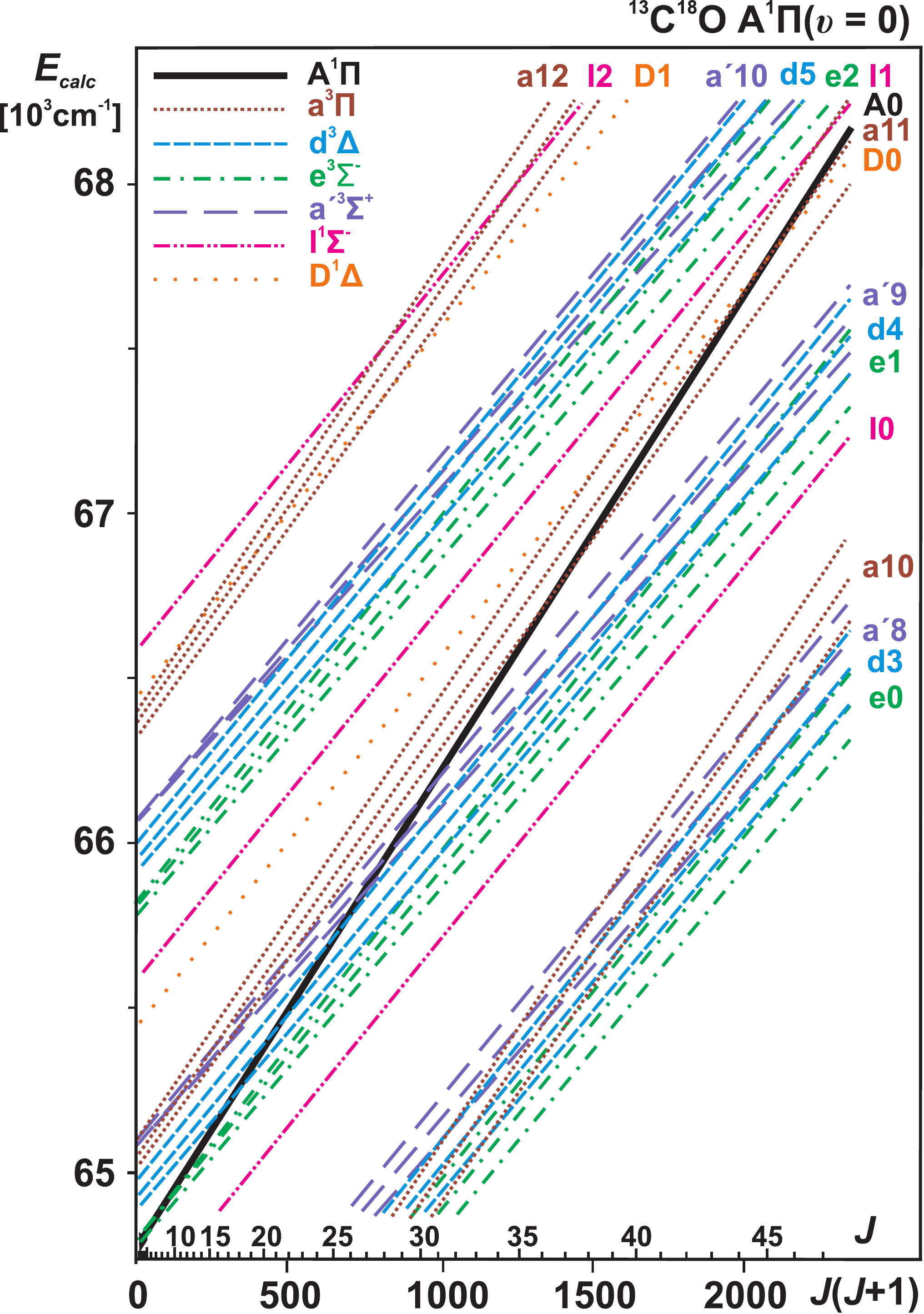}
    \caption{Level diagram for the \CO{} \As(\0) state and its neighbourhood in the region of 65,000 - 68,000 \wn. The labels denote the electronic state, and to their right the vibrational quantum number. The levels were obtained from the mass-scaled equilibrium molecular constants calculated on the basis of Refs. \cite{Niu2013,Niu2016} for \As, Refs. \cite{Kittrell1989,Lefloch1987} for \Ds, Refs. \cite{Field-thesis,Yamamoto1988} for \ats, as well as Refs. \cite{Field-thesis,Lefloch1987} for the \aps, \Is, \ds, and \es ~states. The \CO ~\Xs ~$G$(0) value was taken from Ref. \cite{Coxon2004} to obtain $T_{v=0}$ term.}
    \label{fig:perturbations}
  \end{center}
\end{figure}

\begin{figure}
  \begin{center}
    \includegraphics[width=\textwidth]{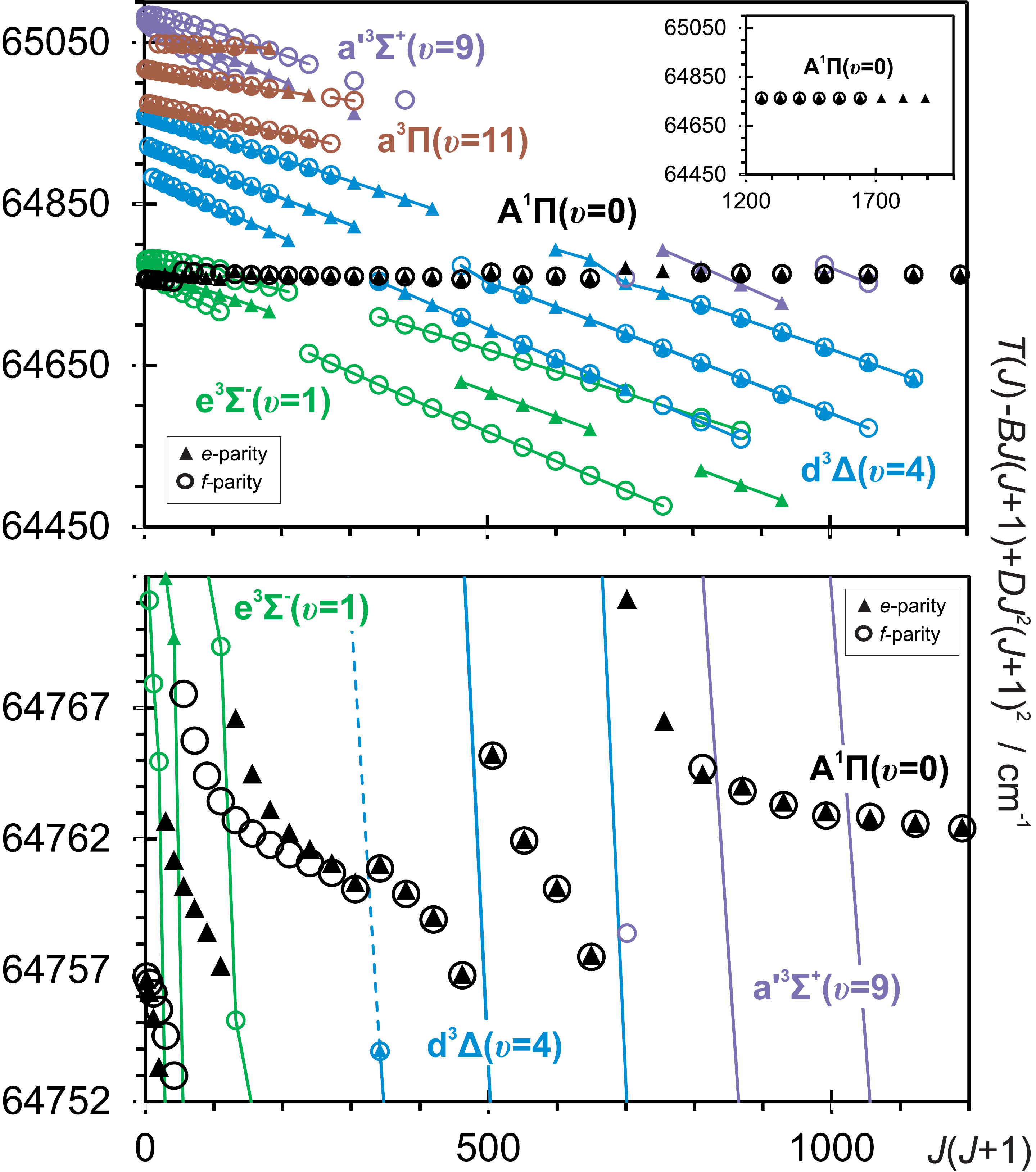}
    \caption{Experimental reduced terms of the \CO{} \As(\0) level and its perturbers. The energies are calculated as $T(J)-BJ(J+1)+DJ^2(J+1)^2$ for $B=1.457$ \wn\ and $D=6.083 \times 10^{-6}$ \wn.}
    \label{fig:ReducedT}
  \end{center}
\end{figure}

It is well known that the \As($v=0$) level is extensively perturbed in all CO isotopologues, with the occurrence of multiple rotational-level crossings with other electronic-vibrational states and smaller effects due to more remote non-crossing levels \cite{Field1972,Niu2013,Hakalla2017}. The long-lived \es$(v=1)$, \ds$(v=4)$, \aps$(v=9)$ levels are primarily responsible for the perturbations through their homogeneous spin-orbit interaction with \As($v=0$), with magnitudes parameterised by $\eta$.
Additionally, the shorter-lived \Ds$(v=0)$ and \Is$(v=0,1)$ states heterogeneously perturb \As($v=0$) through $L$-uncoupling interactions, parameterised by $\xi$.
Rostas and co-workers \cite{Rostas2000,LeFloch1990,Eidelsberg2003} showed that interactions with multiple \As{} vibrational levels contribute to the intensity borrowing of forbidden bands. Therefore the present deperturbation analysis includes some additional levels that affect \As($v=0$) relatively weakly or indirectly, that is \es$(v=0, 2)$, \ds($v=3, 5$), \aps($v=8, 10$), and \ats($v=10, 11$). They were not considered in our work on other CO isotopologues but are included now in the light of a larger experimental data set.
Figure~\ref{fig:perturbations} shows plots of calculated rovibronic level energies against $J(J+1)$ for \As($v=0$) and its nearest neighbours, showing the crossing points where local perturbations may occur, whereas Fig.~\ref{fig:ReducedT} presents experimental reduced terms obtained in this work.

For the mutual interactions between \ats{} and the other triplet states under consideration, there are two perturbation mechanisms in operation: spin-orbit and those that arise from the $\mathbf{B(R^2)}$ term of the rotational Hamiltonian.
Writing the latter as $\mathbf{R = J - L - S}$ we get $\mathbf{J\cdot L}$, $\mathbf{J\cdot S}$, and $\mathbf{L\cdot S}$ terms, respectively $L$-uncoupling, $S$-uncoupling, and spin-electronic coupling.
The $\Delta\Omega = 0$ spin-electronic matrix element is explicitly related to the $\Delta\Omega = 1$ $L$-uncoupling matrix element, because they both consist of an experimentally determined $\langle \Pi|\mathbf{L^+}|\Sigma \rangle$ factor, multiplied by an explicitly known matrix element factor depending only on the spin \cite{Field1972a}.
This means that the perturbation terms derived from the rotational operator have $\Delta\Omega = 0$ and $\Delta\Omega = 1$ matrix elements, the values of which are explicitly locked together. The \ats{} $\sim$ (\Ds, \Is) and \aps {}$\sim$ \es {} interactions arise from spin-orbit interactions. Finally, the \ds{} $\sim$ (\aps, \es) perturbations come from spin-spin interactions represented by the $\varepsilon$ perturbation parameter. Effects of the direct \As($v=0$){} $\sim$ \ats($v=10, 11$) spin-orbit interactions are too weak to be deduced from the data set. This might be ascribed to very small vibrational overlap integrals $\langle v_{\text{A}(0)}|v_{\text{a}(10)} \rangle = 1.4\times10^{-4}$  and $\langle v_{\text{A}(0)}|v_{\text{a}(11)} \rangle = -1.6\times10^{-3}$. Some of the indirect \As{} $\sim$ (\es, \ds, \aps) {}$\sim$ \ats{} perturbations significantly shift the observed \As{} levels and are analysed in detail. All of the close-lying levels taken into consideration intersect A$(v=0)$ in their zero-order approximation except for I$(v=0, 1)$, D$(v=0)$, e$(v=2)$, d$(v=5)$, and a$'(v=10)$, which nevertheless have a noticeable direct influence on A$(v=0)$.

In order to find deperturbed molecular constants for \As($v=0$) we use the PGOPHER software \cite{pgopher} to model this level and all neighbouring perturber levels with an effective-Hamiltonian matrix with diagonal elements composed of deperturbed constants describing each electronic-vibrational level and off-diagonal elements given by the various possible perturbation parameters arising from the spin-orbit, $L$-uncoupling, spin-electronic and spin-spin operators.
The manifold of levels surrounding A$(v=0)$ is combined with unperturbed models of the X$(v=0)$ and B$(v=0)$ levels to simulate transition frequencies for the experimentally-observed bands: \AX$(0,0)$, \BA$(0,0)$, \Be$(0,1)$, \Bd$(0,4)$, \Bap$(0,9)$,  \BX$(0,0)$, \atX$(11,0)$, \apX$(9,0)$, \dX$(4,0)$, and \eX$(1,0)$.
In total, 908 experimental frequencies from 10 bands of \CO{} were used to iteratively refine the free parameters of the effective Hamiltonian model until good general agreement was obtained.

\begin{ThreePartTable}
  \footnotesize

  \begin{TableNotes}

    \item[$a$] All values are in \wn{}.
    \item[$b$] Molecular constants fitted during the model optimisation have uncertainties indicated in parentheses ($1\sigma$, in units of the least significant digit). All other parameters were fixed  during the fitting procedure. The \CO{} ground state level, X$(v=0)$, was fixed to the following constants determined by Coxon and Hajigeorgiou \cite{Coxon2004}: $G_v=\np{1031.055619}$, $B_v=\np{1.746408199}$, $D_v=\np{5.0488146e-6}$, $H_v=\np{4.35471e-12}$, $L_v=\np{-2.4821e-17}$, $M_v=\np{2.99e-23}$, $N_v=\np{-4.5e-28}$, and $O_v=\np{-2.2e-33}$, all values in \wn.
    \item[$c$] Calculated from Ref. \cite{Field-thesis} based on mass-scaling, where the spin - spin constant $\lambda = - 1.5 \times C$ and the $\Lambda$-doubling constants $o = C^{\delta}$ and $q = 2 \times B_{0^+}$. The \CO{} \Xs{} $G(0)$ value was taken from \cite{Coxon2004} to obtain the $T_{v=0}$ term.
    \item[$d$] After Ref. \cite{Lemaire2016}.
    \item[$e$] After Ref. \cite{Prasad1988}.
    \item[$f$] Calculated from Refs. \cite{Niu2013,Niu2016} based on mass-scaling.
    \item[$g$] On the basis of the isotopically recalculated values given by Refs.\cite{Hakalla2017,Hakalla2013} for the \Bs\ as well as using constants determined by Ref. \cite{Coxon2004} for the \Xs\ state.
    \item[$h$] After Ref. \cite{Kepa1988}.
    \item[$i$] After Ref. \cite{Malak1984}.
    \item[$j$] After Ref. \cite{Haridass1994b}.
    \item[$k$] Calculated from Ref. \cite{Kittrell1989} based on mass-scaling.
    \item[$l$] Calculated from Ref. \cite{Lefloch1987} based on mass-scaling.
    \item[$m$] The theoretical spin-orbit $\eta$ and rotational-electronic $\xi$ interaction parameter values were calculated on the basis of the \textbf{a$_{A \sim}$} and \textbf{b$_{A \sim}$} isotopically-invariant given by Hakalla et al. \cite{Hakalla2017} and the appropriate vibrational overlap integrals for \CO\ (see text for details).
    \item[$n$] The value was deduced using elements of an effective Hamiltonian matrix by Field \cite{Field1972} (Table IV) and electronic perturbation matrix elements by Ref. \cite{Field1972a} (Table IV) and compare them with the symmetrised matrix elements implemented in the current fit \cite{pgopher}(see text for details).
    \item[$o$] Determined on the assumption that $\xi \ll \eta$, which condition is very well fulfilled in the present case.
    \item[$p$] The spin-spin off-diagonal interaction.
    \item[$r$] Calculated from Ref. \cite{Field-thesis} based on mass-scaling, where $A_D = 2\times A_J$ \cite{Denijs2011}.
    \item[$s$] Obtained and isotopically recalculated from Ref. \cite{Yamamoto1988}, where the spin-spin constant $\lambda = 1.5 \times \varepsilon$ and the $\Lambda$-doubling constant $p = 2\times p_+$.
  \end{TableNotes}

  \begin{longtable}{l@{\hspace{20pt}}l@{\hspace{20pt}}l@{\hspace{20pt}}l@{\hspace{20pt}}l@{\hspace{20pt}}}
    \caption {Deperturbed molecular parameters for the \As($v=0$) level and its direct and indirect perturbers in \CO\ as well as for the \Bs($v=0$) level.$^{a,b}$ Perturbation parameters as discussed in this work and in Refs.~\cite{Niu2013,Hakalla2016,Trivikram2017}.\label{Molecons}}\\
    \endfirsthead
    \noalign{\ldots{}continued from previous page\hfill}\\
    % \colrule
    \endhead
    % \colrule
    \\\noalign{continued on next page\ldots\hfill}
    \endfoot
    \colrule
    \insertTableNotes
    \endlastfoot

    \colrule
    Constant                 	&		     \As $(v=0)$        		&		      \Bs $(v=0)$     		&		 \Is $(v=0)$            		&		      \Is $(v=1)$          		\\	 \colrule
    $T_v$                    	&	$	64 762.750 18(60)	$	&	$	86 917.360 32(86)	$	&	$	64571.871{}^{c}	$	&	$	65593.173{}^{c}	$	\\	
    &	$	64 763.084{}^{c}	$	&	$	86 916.702{}^{g}	$	&	$		$	&	$		$	\\	
    &	$	64 777.936{}^{d}	$	&	$		$	&	$		$	&	$		$	\\	
    $B$                      	&	$	1.457 416 5(41)	$	&	$	1.769 764 5(30)	$	&	$	1.145 43{}^{c}	$	&	$	1.130 21{}^{c}	$	\\	
    &	$	1.457 4(3){}^{e}	$	&	$	1.769 689(76){}^{h}	$	&	$		$	&	$		$	\\	
    &	$	1.457 46{}^{c}	$	&	$	1.769 653(48){}^{i}	$	&	$		$	&	$		$	\\	
    &	$	1.32{}^{d}	$	&	$	1.769 77(6){}^{j}	$	&	$		$	&	$		$	\\	
    &	$		$	&	$	1.769 7(1){}^{e}	$	&	$		$	&	$		$	\\	
    $D\times 10^{6}$         	&	$	6.083 0(28)	$	&	$	5.538 7(15)	$	&	$	5.65{}^{l}	$	&	$	5.67{}^{l}	$	\\	
    &	$	5.697{}^{c}	$	&	$	5.524(89){}^{h}	$	&	$		$	&	$		$	\\	
    &	$	-194.20{}^{d}	$	&	$	5.49(34){}^{i}	$	&	$		$	&	$		$	\\	
    &	$		$	&	$	6.1(1){}^{j}	$	&	$		$	&	$		$	\\	
    &	$		$	&	$	5.8(2){}^{e}	$	&	$		$	&	$		$	\\	
    $H\times 10^{12}$        	&	$	-12.8{}^{f}	$	&	$		$	&	$	2.25{}^{l}	$	&	$	2.25{}^{l}	$	\\	
    $q\times 10^{5}$        	&	$	2.53(18)	$	&	$		$	&	$		$	&	$		$	\\	
    &	$	-1.19^{f}	$	&	$		$	&	$		$	&	$		$	\\	
    $\xi (\sim$ A, $v=0)$	&	$		$	&	$		$	&	$	-0.032{}^{l}	$	&	$	0.057{}^{m}	$	\\	
    $\eta (\sim$ a, $v=11)$	&	$		$	&	$		$	&	$		$	&	$	-2.409{}^{n}	$	\\
    \pagebreak
    \colrule
    Constant                 	&		     \Ds $(v=0)$          		&		     \es $(v=0)$        		&		     \es $(v=1)$        		&		     \es $(v=2)$        		\\	 \colrule
    $T_v$                    	&	$	65448.421{}^{k}	$	&	$	63 729.173{}^{c}	$	&	$	64 774.962 71(54)	$	&	$	65 802.444{}^{c}	$	\\	
    $B$                      	&	$	1.1339{}^{\rm k}	$	&	$	1.158 49{}^{c}	$	&	$	1.142 611(12)	$	&	$	1.127 38{}^{c}	$	\\	
    $D\times 10^{6}$         	&	$	5.81{}^{k}	$	&	$	5.67{}^{c}	$	&	$	5.571(18)	$	&	$	5.58{}^{c}	$	\\	
    $H\times 10^{12}$        	&	$	-0.22{}^{l}	$	&	$	-1.50{}^{l}	$	&	$	-1.50{}^{l}	$	&	$	-1.50{}^{l}	$	\\	
    $\lambda$                	&	$		$	&	$	0.52{}^{c}	$	&	$	0.536 3(15)	$	&	$	0.54{}^{c}	$	  \\	
    $\gamma\times 10^{3}$    	&	$		$	&	$		$	&	$	-2.401(88)	$	&	$		$	  \\	
    $\eta (\sim$ A, $v=0)$	&	$		$	&	$	-8.480{}^{m}	$	&	$	14.400 1(13)	$	&	$	-17.544{}^{m}	$	  \\	
    &	$		$	&	$		$	&	$	14.143{}^{m}	$	&	$		$	  \\	
    $\xi (\sim$ A, $v=0)$	&	$	0.019{}^{m}	$	&	$		$	&	$		$	&	$		$	  \\	
    $\eta (\sim$ a, $v=11)$	&	$	-0.339{}^{n}	$	&	$		$	&	$	2.516{}^{n}	$	&	$		$	  \\	
    $\xi (\sim$ a, $v=11)$	&	$		$	&	$		$	&	$	-0.056{}^{n}	$	&	$		$	  \\	 \colrule
    Constant                 	&		    \ds $(v=3)$       		&		    \ds $(v=4)$       		&		    \ds $(v=5)$       		&		    \aps $(v=8)$       		\\	 \colrule
    $T_v$                    	&	$	63 886.481{}^{c}	$	&	$	64 928.697 25(84)	$	&	$	65 953.987{}^{c}	$	&	$	64 073.372{}^{c}	$	\\	
    $B$                      	&	$	1.138 29{}^{c}	$	&	$	1.122937(10)	$	&	$	1.108 73{}^{c}	$	&	$	1.093 86{}^{c}	$	\\	
    $D\times 10^{6}$         	&	$	5.361{}^{c}	$	&	$	5.153(12)	$	&	$	5.327{}^{c}	$	&	$	5.19{}^{c}	$	\\	
    $H\times 10^{12}$        	&	$	-0.60{}^{l}	$	&	$	-0.60{}^{l}	$	&	$	-0.60{}^{l}	$	&	$	-0.30{}^{l}	$	\\	
    $A$                      	&	$	-15.649{}^{c}	$	&	$	-16.581 7(19)	$	&	$	-15.909{}^{c}	$	&	$		$	\\	
    $A_D\times 10^{4}$       	&	$	-0.92{}^{r}	$	&	$	-0.92{}^{r}	$	&	$	-0.92{}^{r}	$	&	$		$	  \\	
    $\lambda$                	&	$	0.67{}^{c}	$	&	$	1.124 2(23)	$	&	$	0.85{}^{c}	$	&	$	-1.11{}^{c}	$	  \\	
    $\gamma\times 10^{3}$    	&	$	-4.95{}^{c}	$	&	$	-4.35(18)	$	&	$	-6.28{}^{c}	$	&	$	-5.43{}^{c}	$	  \\	
    $\eta (\sim$ A, $v=0)$	&	$	27.748(42)	$	&	$	-22.133 6(39)	$	&	$	19.764{}^{m}	$	&	$	-4.036{}^{m}	$	  \\	
    &	$	25.7037{}^{m}	$	&	$	-23.391{}^{m}	$	&	$		$	&	$		$	  \\	
    $\eta (\sim$ a, $v=10){}^{o}$	&	$		$	&	$	-32.405{}^{n}	$	&	$		$	&	$		$	  \\	
    $\xi (\sim$ a, $v=10)$	&	$		$	&	$	0.073{}^{n}	$	&	$		$	&	$		$	  \\	
    $\eta (\sim$ a, $v=11){}^{o}$	&	$		$	&	$	-34.503(24)	$	&	$	-33.533{}^{n}	$	&	$	-16.5999{}^{n}	$	  \\	
    &	$		$	&	$	-25.770{}^{n}	$	&	$		$	&	$		$	  \\	
    $\xi (\sim$ a, $v=11)$	&	$		$	&	$	0.0598(17)	$	&	$	0.074{}^{n}	$	&	$	0.039{}^{n}	$	  \\	
    &	$		$	&	$	 0.054{}^{n}	$	&	$		$	&	$		$	  \\	
    $\varepsilon (\sim$ a', $v=9){}^{p}$	&	$		$	&	$	0.196(37)	$	&	$		$	&	$		$	  \\	 \colrule
    Constant                 	&		    \aps $(v=9)$       		&		    \aps $(v=10)$       		&		    \ats $(v=10)$        		&		    \ats $(v=11)$        		\\	 \colrule
    $T_v$                    	&	$	65 078.497 5(41)	$	&	$	66 066.949{}^{c}	$	&	$	63 642.313{}^{c}	$	&	$	65 012.3093(45)	$	\\	
    $B$                      	&	$	1.079 761(32)	$	&	$	1.065 57{}^{c}	$	&	$	1.358 99{}^{c}	$	&	$	1.341 838(54)	$	\\	
    $D\times 10^{6}$         	&	$	5.188(34)	$	&	$	5.17{}^{c}	$	&	$	5.59{}^{c}	$	&	$	5.51(17)	$	\\	
    $H\times 10^{12}$        	&	$	-0.30{}^{l}	$	&	$	-0.30{}^{l}	$	&	$		$	&	$		$	\\	
    $A$                      	&	$		$	&	$		$	&	$	37.50{}^{c}	$	&	$	39.280 9(55)	$	\\	
    $A_D\times 10^{4}$       	&	$		$	&	$		$	&	$	-2.29 {}^{s}	$	&	$	-2.17 {}^{s}	$	  \\	
    $\lambda$                	&	$	-1.144 5(44)	$	&	$	-1.09{}^{c}	$	&	$	-0.0012 {}^{s}	$	&	$	-0.0121(36)	$	  \\	
    $\gamma\times 10^{3}$    	&	$	-7.02(25)	$	&	$	-5.14{}^{c}	$	&	$	3.71{}^{c}	$	&	$	3.53{}^{c} 	$	  \\	
    $o$       	&	$		$	&	$		$	&	$	0.67{}^{c}	$	&	$	0.909 8(33)	$	  \\	
    $p\times 10^{3}$       	&	$		$	&	$		$	&	$	2.91{}^{s}	$	&	$	1.43(52)	$	  \\	
    $q\times 10^{5}$       	&	$		$	&	$		$	&	$	3.27{}^{c}	$	&	$	3.106{}^{c}	$	  \\	
    $\eta (\sim$ A, $v=0)$	&	$	2.739(17)	$	&	$	-1.8865{}^{m}	$	&	$		$	&	$		$	  \\	
    &	$	2.8137{}^{m}	$	&	$		$	&	$		$	&	$		$	  \\	
    $\eta (\sim$ a, $v=10){}^{o}$	&	$	15.1246{}^{n}	$	&	$		$	&	$		$	&	$		$	  \\	
    $\xi (\sim$ a, $v=10)$	&	$	-0.038{}^{n}	$	&	$		$	&	$		$	&	$		$	  \\	
    $\eta (\sim$ a, $v=11){}^{o}$	&	$	3.450(61)	$	&	$		$	&	$		$	&	$		$	  \\	
    &	$	13.195{}^{n}	$	&	$		$	&	$		$	&	$		$	  \\	
    $\xi (\sim$ a, $v=11)$	&	$	-0.0018(10)	$	&	$		$	&	$		$	&	$		$	  \\	
    &	$	 -0.035{}^{n}	$	&	$		$	&	$		$	&	$		$	  \\	
    \colrule                 																		
    \colrule                 																		
    \pagebreak

  \end{longtable}
\end{ThreePartTable}

Full details of our methodology were presented in previous works \cite{Niu2013,Hakalla2016} and the Hamiltonian used in the is described by Western \cite{Western2017}. The explicit formulation of the effective Hamiltonian and matrix elements are contained in the Pgopher file, with a final version is provided in the Supplementary Material. Initial estimates for the parameter values governing excited states are adopted in analogy to other CO isotopologues, using the isotope-scaling constants deduced by Field et al. \cite{Field-thesis,Field1972, Field1972a}, Niu et al. \cite{Niu2013,Niu2016}, Le Floch \cite{Lefloch1987}, Yamamoto et al. \cite{Yamamoto1988}, and Kittrell et al. \cite{Kittrell1989}.
Ground state constants for \CO{} are taken from Coxon and Hajigeorgiou \cite{Coxon2004} and are kept fixed in all fitting procedures.
Constants describing B$(v=0)$ were fit to its experimentally deduced term values.
A computed correlation matrix of all model parameters is monitored during the fitting process to determine a minimal set of molecular constants necessary to model the experimental data.
Some parameters, afflicted by a high degree of correlation with others but verified to be significant were held fixed to estimated values.
They are calculated as described by Hakalla et al.~\cite{Hakalla2016,Hakalla2017}. The value of interactions involving the \ats($v=10, 11)$ levels are calculated using elements of an effective Hamiltonian matrix defined in Field \cite{Field1972} optimised by comparing them with the symmetrised matrix elements used in the current fit \cite{pgopher} as well as electronic perturbation matrix elements given in Table IV of Ref. \cite{Field1972a}. In the final fit, $39$ independent parameters were adjusted and their best-fit values are listed in Table \ref{Molecons}. The root-mean-square error of modelled transition frequencies is then $0.02$ \wn.

All the \ats{} $\sim${} \ds{} and \ats{} $\sim${} \aps{} interactions reported in Table \ref{Molecons} have $\eta$ and $\xi$ parameters with opposite sign. This is a consequence of the dominant electronic configurations involved: the $\Pi$ states have a singly occupied (less than half full) $\pi^{\ast}$ orbital and the $\Sigma$ and $\Delta$ states have a $\pi^3$ (more than half filled) $\pi$ orbital. This means that the two kinds of interaction matrix elements will always have opposite signs for the states of interest in CO.

We find anomalously-small values for the $\eta_{\text{a}(11)\sim \text{a}'(9)} = 3.450(61)$\,\wn\ and $\xi_{\text{a}(11) \sim \text{a}'(9)} = -0.0018(10)$\, \wn\ perturbation parameters (listed in Table~\ref{Molecons}) relative to mass-scaling predictions, yielding $13.195$ \wn\ and $-0.035$ \wn, respectively, as well as a surprisingly large magnitude $\eta_{\text{a}(11)\sim \text{d}(4)} = -34.503(24)$ \wn\ value in comparison with the mass scaling value, $-25.770$ \wn.
The vibrational overlap integrals between the a$(v=11)$ and d$(v=4)$ triplet states can be locally and sensitively affected by a node near the internuclear distance of the crossing between the potential energy curves of the two interacting electronic states, and the interaction parameter deviations are probably related to an imperfect knowledge of the \ats{} potential energy curve employed in the mass-scaling calculations, which is not well characterised above $v=6$ \cite{Yamamoto1988}. However, no similar problem is observed for any of the other isotopologues and its resolution must await clarification by obtaining spectra of the \ats{} state at higher $v$.

Some perturbation mechanisms in addition to spin-orbit interactions ($\eta$) and $L$-uncoupling ($\xi$) were examined but their inclusion in the fit model was found to be statistically unjustified given the accuracy of our experimental ro-vibronic data. Specifically, a second-order spin-spin contribution ($\epsilon_{\Pi\Sigma}$) to the $\Omega = 0$ $\Lambda$-doubling of $^3\Pi$ states (mediated via $\Sigma^+$ and $\Sigma^-$ states) as well as a second-order $\mathbf{H^{SO}} \times \mathbf{H^{ROT}}$ interaction term ($p_3$)~\cite{Effantin1982,Amiot1986a,Wada2000} were considered.

The direct or indirect influence upon A$(v=0)$ of six levels additional to those listed in Table~\ref{Molecons} was tested and ruled out.
These higher- and lower-$v$ vibrational levels of the various electronic states in Table~\ref{Molecons} were found to have either no measurable impact on A$(v=0)$ when included in our model along with estimated interaction parameters, or were highly correlated with molecular constants and/or stronger perturbing effects already included. Details of all the 87 tested interactions are gathered in Table \ref{87interactions}. We believe that the present deperturbation treatment for the A$(v=0)$ state is now limited only by the accuracy and extent of the fitted data set. Perhaps adding more levels and interactions independently quantified with the aid of further spectroscopic measurements would likely expose sensitivity to still-more-remote levels, and it may be that the limits of a reasonable semi-empirical deperturbation treatment has been reached with this analysis.
An existing effective Hamiltonian model of the B$\,{}^3\Sigma^+_u$ and B$''\,{}^3\Pi_u$ states of S$_2$ \cite{Green1996} is more far-reaching in terms of its energy range than what is done here but this system consists of only two electronic states and exhibits many level crossings.

The $\Lambda$-doubling constant of the \As($v=0$) level generally has a small value for the CO isotopologues. The presently determined value of $q = 2.53(18)\times 10^{-5}$ \wn\ has the opposite sign to a value predicted from mass scaling of the main isotopologue: $q = -1.19\times 10^{-5}$ \wn. This $\Lambda$-doubling is, in effect, the result of interactions between many levels in the molecule and its modelled value depends sensitively on which levels are excluded from the effective Hamiltonian matrix. The splitting of $e$- and $f$-parity levels in a $^1\Pi$ state is the result of interactions with states of $\Sigma$ symmetry. The number of $\Sigma$ states explicitly included in the present analysis has increased from our previous work, and a poor extrapolation of the $q$-parameter is then unsurprising.

Rotational-level mixing coefficients and intensity borrowing was also computed using the PGOPHER program for the model \AX$(0,0)$ transitions and its associated extra lines.
Only the unperturbed A$-$X$(0,0)$ transition has a nonzero transition dipole moment and any reduction in its perturbed line strengths is proportional to the fractional admixture of other states into the A$(v=0)$ level.
\begin{figure}
  \includegraphics[width=\textwidth]{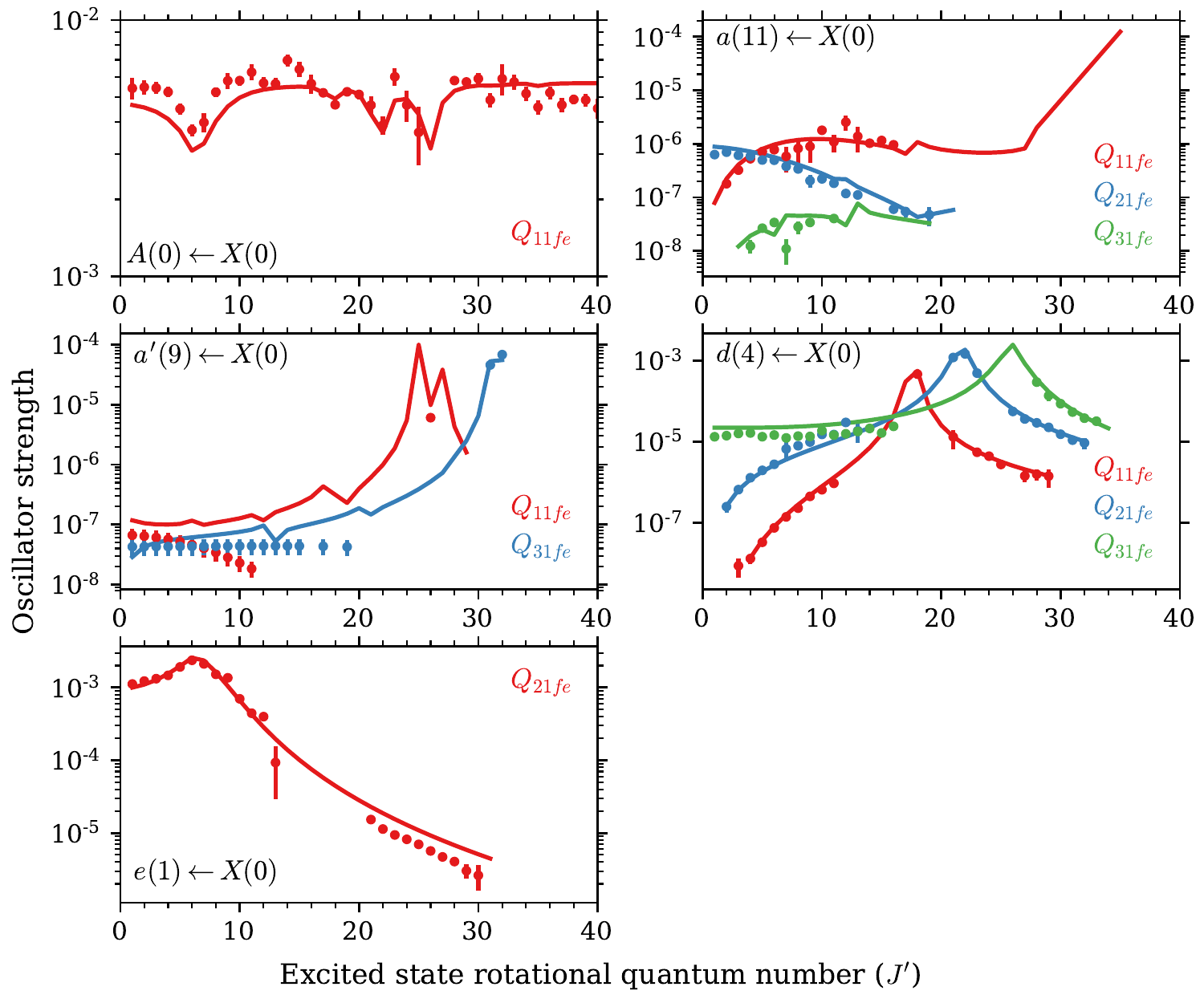}
  \caption{Measured absorption oscillator strengths for Q-branch transitions (points and 1σ error bars indicating random fitting uncertainties) and strengths computed from the effective Hamiltonian model (lines). The model strengths were scaled to match the experimental A(v=0)-X(v=0) data and all plotted data is subject to a common 30\% absolute uncertainty.}
  \label{fig:fvalues}
\end{figure}

Computed and measured oscillator strengths of the \AX$(0,0)$ and forbidden transition $Q$-branches are plotted in Fig.~\ref{fig:fvalues} and show generally-good agreement. All of the $Q$-branch transitions in Fig.~\ref{fig:fvalues} terminate on excited $f$-parity levels.
A qualitatively-similar picture and the same level of agreement between modelled and experimental oscillator strengths is found for $P$- and $R$-branch transitions that terminate on $e$-parity levels.
These strengths have been reduced by factoring out rotational linestrengths for a pure ${}^1\Pi-{}^1\Sigma$ transition, and large dips in A$-$X$(0,0)$ strengths near $J=6$, 21, and 27, are the result of level crossings and increased admixture of the \es$(v=1)$, and \ds$(v=4)$ $F_2$ and $F_3$ states, respectively.
Good agreement is found for the oscillator strengths in the \apX$(11,0)$, \dX$(4,0)$, and \eX$(1,0)$ bands.
A significant disagreement between modelled and measured strengths occurs for the $Q_{11fe}$ branch of \apX$(9,0)$, where the calculated line strengths are significantly larger than observed for $J\gtrsim 5$, while the correct strength is found for the $Q_{11fe}(26)$ line that is most strongly mixed with the A$-$X$(0,0)$ $Q(26)$ transition.
This suggests that the direct spin-orbit interaction of A($v=0$) and a$'(v=9)$ is correctly modelled but that indirect mixing via the a$(v=11)$ intermediary is not completely reproduced in the analysis.
Alternatively, interactions with states not included in the effective Hamiltonian may be involved, or further intensity borrowing from A$-$X$(1,0)$.

\section{Conclusion}
\label{sec:Concl}

The present study focuses on a comprehensive analysis of spectroscopic data for the \As($v=0$) state of the \CO\ isotopologue of the carbon monoxide molecule. It is a member of a sequence of studies analysing \As($v=0$) for the isotopologues $^{12}$C$^{16}$O \cite{Niu2013}, $^{13}$C$^{16}$O \cite{Niu2016b}, $^{12}$C$^{18}$O \cite{Trivikram2017}, and $^{13}$C$^{17}$O \cite{Hakalla2017}. The complementary properties of various state-of-the-art spectroscopic instruments were exploited for gathering a wealth of accurate information from spectral lines connecting a variety of mutually interacting rovibronic states: the extreme absolute accuracy of a $2+1'$ resonance enhanced two-photon laser ionisation study employing Doppler-free excitation in a molecular beam, the photo-emission spectrum from a discharge resolved by visible Fourier-transform spectroscopy, and VUV Fourier transform absorption spectroscopy at the SOLEIL synchrotron. All studies were performed at high resolution and special techniques were employed to access high rotational states: notably high temperature and high pressure. The accuracies of measured transition frequencies for the best lines were respectively $0.002$ \wn, $0.003$ \wn, and $0.02$ \wn\ in the laser-based, visible FT and synchrotron studies.
The level structure of the \As($v=0$) state of \CO {} was targeted via the  \AX$(0, 0)$ and \BA$(0, 0)$ bands, while information was also gathered on the direct and indirect perturber states \es$(v=0,1,2)$, \ds$(v=3,4,5)$, \aps$(v=8,9,10)$, \Ds($v=0$), \Is($v=0,1$), and \ats$(v=10,11)$. The \BX$(0, 0)$ band was investigated by FT-VUV spectroscopy to connect the visible emission study on an absolute energy scale with respect to the ground state. Weak perturbations in the \Bs(\0) level were observed and included in the analysis.

A comprehensive set of deperturbed constants and level energies for the A($v=0$) state and its perturbers is determined from this set of combined data. The complexity of the present deperturbation analysis exceeds that of the previous studies on other isotopologues and is made possible here by a more extensive highly-accurate data set. The number of modelled perturber states is extended and we determine molecular constants for some triplet levels that do not exhibit a crossing with \As($v=0$). In general, up to 87 vibronic interactions are considered  between A($v=0$), A($v=1$), I($v=0,1,2$) and D($v=0,1,2$) singlet states and a large set of the a($v=10,11,12$), d($v=3,4,5$), a$'$($v=8,9,10$), and e($v=0,1,2,3$) triplet levels. Some mutual interactions of various triplet states perturbing the A($v=0$) level are determined in the analysis which had not been distinguishable in our previous studies focusing on other isotopologues. Although eight lines of the vast body of observed transitions could not be identified, there is no need to invoke additional band systems beyond those well-known to CO to describe the observed spectroscopic patterns, as was done in Ref.~\cite{Lemaire2016}.

The present study surpasses in complexity any previous similar analyses of the CO molecule, which is a prototypical species for perturbations. There are 15 mutually interacting electronic-vibrational levels included to reproduce over 900 line frequencies to a high level of accuracy. The deperturbation model also reproduces the borrowing of absorption oscillator strength by the observed forbidden bands from the main transition \AX$(0,0)$.

\section*{Acknowledgements}

The authors are grateful to the general and technical staff of SOLEIL
for providing beam time under projects no. 20120653 and no.
20160118. RH thanks LASERLAB-EUROPE for support of this research
(Grants EUH2020-RIP-654148 and EC's-SPF-284464) as well as European
Regional Development Fund and the Polish state budget within the
framework of the Carpathian Regional Operational Programme
(RPPK.01.03.00-18-001/10) through the funding of the Center for
Innovation and Transfer of Natural Science and Engineering Knowledge
of the University of Rzesz\'{o}w. AH acknowledges support from the
postdoctoral fellowship program of PSL Research University Paris and
NASA Postdoctoral Program. WU acknowledges financial support from the
European Research Council (ERC) under the European Union's Horizon
2020 research and innovation programme (Grant Agreement No. 670168).
RWF is grateful to the US National Science Foundation (Grant CHE-1361865).

\section*{Supplementary Material}
\label{sec:Suppl}

In the Supplementary material details of the spectroscopic results on the \CO{} molecule are presented in the form of Tables as well as electronic data files contain Pgopher final fit and line lists used in the global analysis.
Table \ref{A0-term} presents the term values for the \As($v=0$) state, Table \ref{d4-term} term values for \ds($v=4$), Table \ref{e1-term} for the \es($v=1$) state, Table \ref{a11-term} for the \ats($v=11$) state, and Table \ref{Table-ap9} for the \aps($v=9$) state. Table \ref{87interactions} highlights details of all 87 interactions tested in the framework of the \CO{} deperturbation analysis, while in Table \ref{B-term} the term values of the B($v=0$) state are presented.

% \bibliography{CO}
%\bibliographystyle{amsplain}
%\bibliographystyle{plain}
%\bibliographystyle{tfo}
%\bibliographystyle{abbrvnat}
%\bibliographystyle{tMPH}
% \bibliographystyle{apsrev4-1}
%\bibliographystyle{unsrt}

\clearpage

\section*{Supplementary Material}

\clearpage

\begin{table*}
\caption{Term values of $A(0)$\textsuperscript{$a$}.}
\footnotesize
\begin{threeparttable}
\begin{tabular}{cll}
\label{A0-term}
\colrule
$J$                & \multicolumn{1}{c}{$F_{1e}$}           & \multicolumn{1}{c}{$F_{1f}$}           \\
\colrule
  1                & 64\,759.580(1)     & 64\,759.677(1)     \\
  2                & 64\,764.914(1)     & 64\,765.268(1)     \\
  3                & 64\,772.6806(9)    & 64\,773.608(1)     \\
  4                & 64\,782.467(1)     & 64\,784.635(1)     \\
  5                & 64\,806.4001(9)    & 64\,798.220(1)     \\
  6                & 64\,822.406(1)     & 64\,814.185(1)     \\
  7                & 64\,841.800(1)     & 64\,849.114(1)     \\
  8                & 64\,864.272(1)     & 64\,870.642(1)     \\
  9                & 64\,889.570(1)     & 64\,895.512(2)     \\
 10                & 64\,917.404(1)     & 64\,923.663(1)     \\
 11                & 64\,958.850(1)     & 64\,954.966(3)     \\
 12                & 64\,991.683(2)     & 64\,989.39(4)      \\
 13                & 65\,028.155(1)     & 65\,026.801(1)     \\
 14                & 65\,068.012(2)     & 65\,067.176(2)     \\
 15                & 65\,111.017(2)     & 65\,110.476(2)     \\
 16                & 65\,157.007(3)     & 65\,156.640(2)     \\
 17                & 65\,205.697(2)     & 65\,205.442(2)     \\
 18                & 65\,258.719(4)     & 65\,258.553(3)     \\
 19                & 65\,312.929(4)     & 65\,312.786(3)     \\
 20                & 65\,370.022(7)     & 65\,369.915(4)     \\
 21                & 65\,428.823(7)     & 65\,428.768(7)     \\
 22                & 65\,501.05(1)      & 65\,500.996(9)     \\
 23                & 65\,564.55(2)      & 65\,564.498(8)     \\
 24                & 65\,632.32(2)      & 65\,632.27(1)      \\
 25                & 65\,702.24(2)      & 65\,702.16(3)      \\
 26                & 65\,791.15(5)      & \multicolumn{1}{c}{--}                 \\
 27                & 65\,864.72(5)      & \multicolumn{1}{c}{--}                 \\
 28                & 65\,943.76(5)      & 65\,944.00(5)      \\
 29                & 66\,027.26(5)      & 66\,027.05(5)      \\
 30                & 66\,113.42(5)      & 66\,113.29(5)      \\
 31                & 66\,202.68(5)      & 66\,202.52(5)      \\
 32                & 66\,294.89(5)      & 66\,294.94(5)      \\
 33                & 66\,390.01(5)      & 66\,389.98(5)      \\
 34                & 66\,487.987(5)     & 66\,487.94(5)      \\
 35                & 66\,588.763(5)     & 66\,588.75(5)      \\
 36                & 66\,692.588(5)     & 66\,692.54(5)      \\
 37                & 66\,799.079(5)     & 66\,799.018(5)     \\
 38                & 66\,908.416(5)     & 66\,908.345(5)     \\
 39                & 67\,020.564(6)     & 67\,020.487(5)     \\
 40                & 67\,135.52(5)      & 67\,135.435(9)     \\
 41                & \multicolumn{1}{c}{--}                 & 67\,253.18(2)      \\
 42                & \multicolumn{1}{c}{--}                 & 67\,373.72(3)      \\
 43                & \multicolumn{1}{c}{--}                 & 67\,497.04(5)      \\
\colrule
\end{tabular}
\begin{tablenotes}
\begin{footnotesize}
\item[$a$] In units of \wn{} and with $1\sigma$ fitting uncertainties given in parentheses in units of the least-significant digit that are additional to a 0.01\,\wn{} systematic uncertainty.
\end{footnotesize}
\end{tablenotes}
\end{threeparttable}
\end{table*}

\begin{table*}
\caption{Term values of $d(4)$\textsuperscript{$a$}.}
\footnotesize
\begin{threeparttable}
\begin{tabular}{cllllll}
\label{d4-term}
\colrule
$J$                & \multicolumn{1}{c}{$F_{1e}$}           & \multicolumn{1}{c}{$F_{1f}$}           & \multicolumn{1}{c}{$F_{2e}$}           & \multicolumn{1}{c}{$F_{2f}$}           & \multicolumn{1}{c}{$F_{3e}$}           & \multicolumn{1}{c}{$F_{3f}$}           \\
\colrule
  1                & \multicolumn{1}{c}{--}                 & \multicolumn{1}{c}{--}                 & \multicolumn{1}{c}{--}                 & \multicolumn{1}{c}{--}                 & 64\,962.129(5)     & 64\,962.134(5)     \\
  2                & \multicolumn{1}{c}{--}                 & \multicolumn{1}{c}{--}                 & 64\,929.988(5)     & 64\,930.00(2)      & 64\,966.878(5)     & 64\,966.888(5)     \\
  3                & \multicolumn{1}{c}{--}                 & 64\,900.34(5)      & 64\,936.739(5)     & 64\,936.732(5)     & 64\,974.002(5)     & 64\,974.035(7)     \\
  4                & 64\,908.83(5)      & 64\,908.78(3)      & 64\,945.750(5)     & 64\,945.752(5)     & 64\,983.520(5)     & 64\,983.56(1)      \\
  5                & 64\,919.42(2)      & 64\,919.35(1)      & 64\,957.008(5)     & 64\,957.011(5)     & 64\,995.428(5)     & 64\,995.476(5)     \\
  6                & 64\,932.002(8)     & 64\,932.043(9)     & 64\,970.521(5)     & 64\,970.523(5)     & 65\,009.718(5)     & 65\,009.774(5)     \\
  7                & 64\,946.831(5)     & 64\,946.834(5)     & 64\,986.284(5)     & 64\,986.31(5)      & 65\,026.385(9)     & 65\,026.453(5)     \\
  8                & 64\,963.784(5)     & 64\,963.786(5)     & 65\,004.303(5)     & 65\,004.299(5)     & 65\,045.458(5)     & 65\,045.514(9)     \\
  9                & 64\,982.887(5)     & 64\,982.889(9)     & 65\,024.562(5)     & 65\,024.565(5)     & 65\,066.877(5)     & 65\,066.930(5)     \\
 10                & 65\,004.14(5)      & 65\,004.139(5)     & 65\,047.075(5)     & 65\,047.074(5)     & 65\,090.649(5)     & 65\,090.677(7)     \\
 11                & 65\,027.555(5)     & 65\,027.557(5)     & 65\,071.839(5)     & 65\,071.838(5)     & 65\,116.763(5)     & 65\,116.806(5)     \\
 12                & 65\,053.140(5)     & \multicolumn{1}{c}{--}                 & 65\,098.841(5)     & 65\,098.86(1)      & 65\,145.194(5)     & 65\,145.234(5)     \\
 13                & 65\,080.900(5)     & \multicolumn{1}{c}{--}                 & 65\,128.097(5)     & 65\,128.10(3)      & 65\,175.936(5)     & 65\,175.970(5)     \\
 14                & 65\,110.79(5)      & \multicolumn{1}{c}{--}                 & 65\,159.604(5)     & \multicolumn{1}{c}{--}                 & 65\,208.980(5)     & 65\,209.004(5)     \\
 15                & \multicolumn{1}{c}{--}                 & \multicolumn{1}{c}{--}                 & 65\,193.34(1)      & \multicolumn{1}{c}{--}                 & 65\,244.310(5)     & 65\,244.30(1)      \\
 16                & \multicolumn{1}{c}{--}                 & \multicolumn{1}{c}{--}                 & 65\,229.36(2)      & \multicolumn{1}{c}{--}                 & 65\,281.925(5)     & 65\,281.942(5)     \\
 17                & \multicolumn{1}{c}{--}                 & \multicolumn{1}{c}{--}                 & 65\,267.64(5)      & \multicolumn{1}{c}{--}                 & 65\,321.808(7)     & \multicolumn{1}{c}{--}                 \\
 18                & 65\,251.63(2)      & 65\,251.59(2)      & \multicolumn{1}{c}{--}                 & \multicolumn{1}{c}{--}                 & 65\,363.994(5)     & \multicolumn{1}{c}{--}                 \\
 19                & 65\,292.895(5)     & \multicolumn{1}{c}{--}                 & \multicolumn{1}{c}{--}                 & \multicolumn{1}{c}{--}                 & 65\,408.414(6)     & \multicolumn{1}{c}{--}                 \\
 20                & 65\,336.020(5)     & \multicolumn{1}{c}{--}                 & \multicolumn{1}{c}{--}                 & \multicolumn{1}{c}{--}                 & 65\,455.114(6)     & \multicolumn{1}{c}{--}                 \\
 21                & 65\,381.29(3)      & 65\,381.29(2)      & \multicolumn{1}{c}{--}                 & 65\,445.50(2)      & \multicolumn{1}{c}{--}                 & \multicolumn{1}{c}{--}                 \\
 22                & 65\,428.705(5)     & \multicolumn{1}{c}{--}                 & 65\,486.021(7)     & 65\,485.97(2)      & \multicolumn{1}{c}{--}                 & \multicolumn{1}{c}{--}                 \\
 23                & 65\,478.341(6)     & 65\,478.338(5)     & 65\,540.152(5)     & 65\,540.147(7)     & \multicolumn{1}{c}{--}                 & \multicolumn{1}{c}{--}                 \\
 24                & 65\,530.147(9)     & 65\,530.146(5)     & 65\,594.704(5)     & \multicolumn{1}{c}{--}                 & 65\,665.677(5)     & \multicolumn{1}{c}{--}                 \\
 25                & 65\,584.14(2)      & 65\,584.137(7)     & 65\,650.983(5)     & \multicolumn{1}{c}{--}                 & 65\,725.522(5)     & \multicolumn{1}{c}{--}                 \\
 26                & 65\,640.32(2)      & \multicolumn{1}{c}{--}                 & 65\,709.290(5)     & 65\,709.288(5)     & 65\,771.622(5)     & \multicolumn{1}{c}{--}                 \\
 27                & \multicolumn{1}{c}{--}                 & 65\,698.68(3)      & 65\,769.716(5)     & 65\,769.713(5)     & 65\,838.224(5)     & \multicolumn{1}{c}{--}                 \\
 28                & \multicolumn{1}{c}{--}                 & 65\,759.24(3)      & 65\,832.288(5)     & 65\,832.291(5)     & 65\,904.279(5)     & 65\,904.296(5)     \\
 29                & \multicolumn{1}{c}{--}                 & 65\,821.97(5)      & 65\,897.028(6)     & 65\,897.020(6)     & 65\,971.634(5)     & 65\,971.64(1)      \\
 30                & \multicolumn{1}{c}{--}                 & \multicolumn{1}{c}{--}                 & 65\,963.93(1)      & 65\,963.935(9)     & 66\,040.843(5)     & 66\,040.843(5)     \\
 31                & \multicolumn{1}{c}{--}                 & \multicolumn{1}{c}{--}                 & 66\,033.02(3)      & 66\,033.03(2)      & 66\,112.071(6)     & 66\,112.078(5)     \\
 32                & \multicolumn{1}{c}{--}                 & \multicolumn{1}{c}{--}                 & \multicolumn{1}{c}{--}                 & 66\,104.29(3)      & 66\,185.41(2)      & 66\,185.42(1)      \\
 33                & \multicolumn{1}{c}{--}                 & \multicolumn{1}{c}{--}                 & \multicolumn{1}{c}{--}                 & \multicolumn{1}{c}{--}                 & 66\,260.90(2)      & 66\,260.90(2)      \\
\colrule
\end{tabular}
\begin{tablenotes}
\begin{footnotesize}
\item[$a$] In units of \wn{} and with $1\sigma$ fitting uncertainties given in parentheses in units of the least-significant digit that are additional to a 0.01\,\wn{} systematic uncertainty.
\end{footnotesize}
\end{tablenotes}
\end{threeparttable}
\end{table*}

\begin{table*}
\caption{Term values of $e(1)$\textsuperscript{$a$}.}
\footnotesize
\begin{threeparttable}
\begin{tabular}{clll}
\label{e1-term}
\colrule
$J$                & \multicolumn{1}{c}{$F_{1e}$}           & \multicolumn{1}{c}{$F_{2f}$}           & \multicolumn{1}{c}{$F_{3e}$}           \\
\colrule
  1                & 64\,776.99(2)      & 64\,781.461(7)     & 64\,783.37(1)      \\
  2                & 64\,779.840(7)     & 64\,786.251(4)     & 64\,790.130(9)     \\
  3                & 64\,785.411(3)     & 64\,793.486(5)     & 64\,799.244(8)     \\
  4                & 64\,794.094(2)     & 64\,803.259(5)     & 64\,810.681(8)     \\
  5                & 64\,793.791(4)     & 64\,815.649(2)     & 64\,824.474(9)     \\
  6                & 64\,806.537(4)     & 64\,830.852(2)     & 64\,840.598(5)     \\
  7                & 64\,820.990(7)     & 64\,832.282(2)     & 64\,859.087(4)     \\
  8                & 64\,837.44(2)      & 64\,852.298(2)     & 64\,879.992(3)     \\
  9                & 64\,855.98(1)      & 64\,874.1(1)       & 64\,903.408(3)     \\
 10                & 64\,876.76(3)      & 64\,897.908(4)     & 64\,929.563(3)     \\
 11                & \multicolumn{1}{c}{--}                 & 64\,923.676(7)     & 64\,947.347(2)     \\
 12                & \multicolumn{1}{c}{--}                 & 64\,951.527(9)     & 64\,978.933(3)     \\
 13                & \multicolumn{1}{c}{--}                 & 64\,981.51(7)      & 65\,012.064(8)     \\
 14                & \multicolumn{1}{c}{--}                 & \multicolumn{1}{c}{--}                 & 65\,046.98(1)      \\
 15                & 65\,014.071(5)     & \multicolumn{1}{c}{--}                 & \multicolumn{1}{c}{--}                 \\
 16                & 65\,048.32(1)      & \multicolumn{1}{c}{--}                 & \multicolumn{1}{c}{--}                 \\
 17                & 65\,084.816(5)     & \multicolumn{1}{c}{--}                 & \multicolumn{1}{c}{--}                 \\
 18                & 65\,123.573(5)     & \multicolumn{1}{c}{--}                 & 65\,207.535(5)     \\
 19                & 65\,164.608(5)     & \multicolumn{1}{c}{--}                 & 65\,253.14(3)      \\
 20                & 65\,207.893(5)     & \multicolumn{1}{c}{--}                 & 65\,301.02(5)      \\
 21                & 65\,253.435(5)     & 65\,301.480(5)     & 65\,351.135(5)     \\
 22                & 65\,301.229(5)     & 65\,351.571(5)     & 65\,403.479(5)     \\
 23                & 65\,351.279(7)     & 65\,403.903(5)     & 65\,458.067(5)     \\
 24                & 65\,403.57(1)      & 65\,458.475(5)     & 65\,514.885(5)     \\
 25                & 65\,458.11(2)      & 65\,515.291(5)     & 65\,573.941(5)     \\
 26                & 65\,514.91(2)      & 65\,574.340(5)     & 65\,635.225(8)     \\
 27                & 65\,573.96(3)      & 65\,635.635(8)     & 65\,698.60(2)      \\
 28                & \multicolumn{1}{c}{--}                 & 65\,699.15(1)      & 65\,764.50(1)      \\
 29                & \multicolumn{1}{c}{--}                 & 65\,764.90(2)      & 65\,832.45(5)      \\
 30                & \multicolumn{1}{c}{--}                 & 65\,832.89(4)      & \multicolumn{1}{c}{--}                 \\
\colrule
\end{tabular}
\begin{tablenotes}
\begin{footnotesize}
\item[$a$] In units of \wn{} and with $1\sigma$ fitting uncertainties given in parentheses in units of the least-significant digit that are additional to a 0.01\,\wn{} systematic uncertainty.
\end{footnotesize}
\end{tablenotes}
\end{threeparttable}
\end{table*}

\begin{table*}
\caption{Term values of $a(11)$\textsuperscript{$a$}.}
\footnotesize
\begin{threeparttable}
\begin{tabular}{cllllll}
\label{a11-term}
\colrule
$J$                & \multicolumn{1}{c}{$F_{1e}$}           & \multicolumn{1}{c}{$F_{1f}$}           & \multicolumn{1}{c}{$F_{2e}$}           & \multicolumn{1}{c}{$F_{2f}$}           & \multicolumn{1}{c}{$F_{3e}$}           & \multicolumn{1}{c}{$F_{3f}$}           \\
\colrule
  1                & 64\,976.03(2)      & \multicolumn{1}{c}{--}                 & 65\,019.973(5)     & 65\,019.969(5)     & \multicolumn{1}{c}{--}                 & \multicolumn{1}{c}{--}                 \\
  2                & 64\,981.170(5)     & 64\,983.001(5)     & 65\,025.167(5)     & 65\,025.170(5)     & \multicolumn{1}{c}{--}                 & \multicolumn{1}{c}{--}                 \\
  3                & 64\,988.868(5)     & 64\,990.669(5)     & 65\,032.945(5)     & 65\,032.986(5)     & \multicolumn{1}{c}{--}                 & \multicolumn{1}{c}{--}                 \\
  4                & 64\,999.105(5)     & 65\,000.879(5)     & 65\,043.359(5)     & 65\,043.429(5)     & \multicolumn{1}{c}{--}                 & 65\,077.61(3)      \\
  5                & 65\,011.881(5)     & 65\,013.608(5)     & 65\,056.417(5)     & 65\,056.511(5)     & \multicolumn{1}{c}{--}                 & 65\,092.01(1)      \\
  6                & 65\,027.176(5)     & 65\,028.857(5)     & 65\,072.129(5)     & 65\,072.267(5)     & 65\,109.23(2)      & 65\,109.22(1)      \\
  7                & 65\,044.990(5)     & 65\,046.65(5)      & 65\,090.516(5)     & 65\,090.688(5)     & 65\,129.29(2)      & 65\,129.32(5)      \\
  8                & 65\,065.311(5)     & 65\,066.88(5)      & 65\,111.588(5)     & 65\,111.798(5)     & 65\,152.06(1)      & 65\,152.19(2)      \\
  9                & 65\,088.158(5)     & 65\,089.69(3)      & 65\,135.351(5)     & 65\,135.62(1)      & 65\,177.91(4)      & 65\,177.80(1)      \\
 10                & 65\,113.543(5)     & 65\,115.025(5)     & 65\,161.800(5)     & 65\,162.107(5)     & 65\,206.23(3)      & \multicolumn{1}{c}{--}                 \\
 11                & 65\,141.460(5)     & 65\,142.92(1)      & 65\,190.964(6)     & 65\,191.310(5)     & 65\,237.40(1)      & 65\,237.41(1)      \\
 12                & 65\,171.942(5)     & 65\,173.324(9)     & 65\,222.809(5)     & 65\,223.243(6)     & 65\,271.36(2)      & \multicolumn{1}{c}{--}                 \\
 13                & 65\,204.994(5)     & 65\,206.33(5)      & 65\,257.359(6)     & 65\,257.817(5)     & 65\,308.03(5)      & \multicolumn{1}{c}{--}                 \\
 14                & 65\,240.621(5)     & 65\,241.909(7)     & 65\,294.567(9)     & \multicolumn{1}{c}{--}                 & \multicolumn{1}{c}{--}                 & \multicolumn{1}{c}{--}                 \\
 15                & 65\,278.840(5)     & 65\,280.051(7)     & 65\,334.25(2)      & \multicolumn{1}{c}{--}                 & \multicolumn{1}{c}{--}                 & \multicolumn{1}{c}{--}                 \\
 16                & \multicolumn{1}{c}{--}                 & 65\,320.776(5)     & \multicolumn{1}{c}{--}                 & 65\,377.75(2)      & \multicolumn{1}{c}{--}                 & \multicolumn{1}{c}{--}                 \\
 17                & \multicolumn{1}{c}{--}                 & \multicolumn{1}{c}{--}                 & \multicolumn{1}{c}{--}                 & 65\,423.13(2)      & \multicolumn{1}{c}{--}                 & \multicolumn{1}{c}{--}                 \\
 18                & \multicolumn{1}{c}{--}                 & \multicolumn{1}{c}{--}                 & \multicolumn{1}{c}{--}                 & \multicolumn{1}{c}{--}      & \multicolumn{1}{c}{--}                 & \multicolumn{1}{c}{--}                 \\
 19                & \multicolumn{1}{c}{--}                 & \multicolumn{1}{c}{--}                 & \multicolumn{1}{c}{--}                 & 65\,521.88(4)      & \multicolumn{1}{c}{--}                 & \multicolumn{1}{c}{--}                 \\
\colrule
\end{tabular}
\begin{tablenotes}
\begin{footnotesize}
\item[$a$] In units of \wn{} and with $1\sigma$ fitting uncertainties given in parentheses in units of the least-significant digit that are additional to a 0.01\,\wn{} systematic uncertainty.
\end{footnotesize}
\end{tablenotes}
\end{threeparttable}
\end{table*}

\begin{table*}
\caption{Term values of $a'(9)$\textsuperscript{$a$}.}
\footnotesize
\begin{threeparttable}
\begin{tabular}{clll}
\label{Table-ap9}
\colrule
$J$                & \multicolumn{1}{c}{$F_{1f}$}           & \multicolumn{1}{c}{$F_{2e}$}           & \multicolumn{1}{c}{$F_{3f}$}           \\
\colrule
  1                & 65\,078.40(1)      & 65\,079.97(1)      & 65\,085.99(2)      \\
  2                & 65\,080.77(1)      & 65\,084.287(8)     & 65\,092.30(1)      \\
  3                & 65\,085.180(8)     & 65\,090.767(7)     & 65\,100.84(1)      \\
  4                & 65\,091.694(9)     & 65\,099.403(6)     & 65\,111.587(9)     \\
  5                & 65\,100.377(8)     & 65\,110.202(5)     & 65\,124.493(8)     \\
  6                & 65\,111.237(7)     & 65\,123.164(5)     & 65\,139.593(8)     \\
  7                & 65\,124.14(3)      & 65\,138.286(6)     & 65\,156.849(8)     \\
  8                & 65\,139.26(1)      & 65\,155.620(6)     & 65\,176.267(8)     \\
  9                & 65\,156.54(1)      & 65\,174.81(1)      & 65\,197.846(9)     \\
 10                & 65\,175.97(1)      & 65\,196.493(9)     & 65\,221.594(8)     \\
 11                & 65\,197.58(2)      & 65\,220.24(2)      & 65\,247.50(1)      \\
 12                & \multicolumn{1}{c}{--}                 & 65\,246.09(1)      & 65\,275.60(2)      \\
 13                & \multicolumn{1}{c}{--}                 & 65\,274.20(1)      & 65\,305.48(1)      \\
 14                & \multicolumn{1}{c}{--}                 & 65\,304.39(2)      & 65\,337.96(2)      \\
 15                & \multicolumn{1}{c}{--}                 & \multicolumn{1}{c}{--}                 & 65\,372.44(2)      \\
 17                & \multicolumn{1}{c}{--}                 & 65\,407.68(5)      & 65\,447.81(2)      \\
 19                & \multicolumn{1}{c}{--}                 & \multicolumn{1}{c}{--}                 & 65\,531.81(5)      \\
 26                & 65\,778.418(7)     & \multicolumn{1}{c}{--}                 & \multicolumn{1}{c}{--}                 \\
 27                & \multicolumn{1}{c}{--}                 & 65\,891.17(1)      & \multicolumn{1}{c}{--}                 \\
 28                & \multicolumn{1}{c}{--}                 & 65\,951.394(5)     & \multicolumn{1}{c}{--}                 \\
 29                & \multicolumn{1}{c}{--}                 & 66\,013.077(5)     & \multicolumn{1}{c}{--}                 \\
 30                & \multicolumn{1}{c}{--}                 & 66\,077.39(2)      & \multicolumn{1}{c}{--}                 \\
 31                & \multicolumn{1}{c}{--}                 & \multicolumn{1}{c}{--}                 & 66\,213.693(7)     \\
 32                & \multicolumn{1}{c}{--}                 & \multicolumn{1}{c}{--}                 & 66\,284.016(7)     \\
\colrule
\end{tabular}
\begin{tablenotes}
\begin{footnotesize}
\item[$a$] In units of \wn{} and with $1\sigma$ fitting uncertainties given in parentheses in units of the least-significant digit that are additional to a 0.01\,\wn{} systematic uncertainty.
\end{footnotesize}
\end{tablenotes}
\end{threeparttable}
\end{table*}

\clearpage

\begin{ThreePartTable}
  \footnotesize
  \begin{TableNotes}
    % \begin{footnotesize}
    \item[$a$] Noticeable influence on the constant, interaction (within one standard deviation) and/or residual (within accuracy of the experimental lines) values used in the framework of the final deperturbation fit. It is tested via verifying of the result differences of the appropriate quantities using floated or fixed (to the calculated value and then to 0) interaction parameter.
    % \end{footnotesize}
  \end{TableNotes}
  % \begin{longtable}{c@{\hspace{3pt}}r@{\hspace{3pt}}c@{\hspace{1pt}}l@{\hspace{3pt}}l@{\hspace{3pt}}c@{\hspace{3pt}}c@{\hspace{10pt}}l@{\hspace{20pt}}}
  \begin{longtable}{c@{\hspace{3pt}}r@{\hspace{3pt}}c@{\hspace{1pt}}l@{\hspace{3pt}}l@{\hspace{3pt}}c@{\hspace{3pt}}c@{\hspace{10pt}}p{0.5\textwidth}}
    \caption{All interactions tested in the \CO{} global deperturbation fit.\label{87interactions}}\\
    \endfirsthead
    \noalign{\ldots{}continued from previous page\hfill}\\
    % \colrule
    \endhead
    % \colrule
    \\\noalign{continued on next page\ldots\hfill}
    \endfoot
    \colrule
    \insertTableNotes
    \endlastfoot

    \colrule															
    $N^{o}$	&	Interaction 	&		&		&	Significance\textsuperscript{$a$}	&	Included	&	Floated	&	Notes	\\
    &		&		&		&		&	in the fit	&	/Fixed	&		\\
    \colrule	&		&		&		&		&		&		&		\\
    1	&	 A($v=0$)	&	$\sim$	&	 d($v=3$)	&	Significant	&	Yes	&	Floated	&		\\
    2	&		&	$\sim$	&	 d($v=4$)	&	Significant	&	Yes	&	Floated	&		\\
    3	&		&	$\sim$	&	 d($v=5$)	&	Significant	&	Yes	&	Fixed	&	Strongly correlated with the $B$ constant of A($v=0$).	\\
    4	&		&	$\sim$	&	 e($v=0$)	&	Significant	&	Yes	&	Fixed	&	Strongly correlated with the $B$ and $D$ constants of A($v=0$)	\\
    5	&		&	$\sim$	&	 e($v=1$)	&	Significant	&	Yes	&	Floated	&		\\
    6	&		&	$\sim$	&	 e($v=2$)	&	Significant	&	Yes	&	Fixed	&	Strongly correlated with the $B$ and $D$ constants of A($v=0$) and totally correlated with the A($v=0$){}$\sim$ d($v=3$) interaction.\\
    % &		&		&		&		&		&		&	and totally correlated with the A($v=0$){}$\sim$ d($v=3$) interaction.	\\
    7	&		&	$\sim$	&	 a$'$($v=8$)	&	Significant	&	Yes	&	Fixed	&	Totally correlated with A($v=0$){}$\sim$ d($v=3$).	\\
    8	&		&	$\sim$	&	 a$'$($v=9$)	&	Significant	&	Yes	&	Floated	&		\\
    9	&		&	$\sim$	&	 a$'$($v=10$)	&	Significant	&	Yes	&	Fixed	&	Strongly correlated with the $B$ and $D$ constants of A($v=0$).	\\
    10	&		&	$\sim$	&	 D($v=0$)	&	Significant	&	Yes	&	Fixed	&	Statistically indeterminable. Floating causes divergency of the fit.	\\
    11	&		&	$\sim$	&	 D($v=1$)	&	Insignificant	&	No	&	-	&		\\
    12	&		&	$\sim$	&	 I($v=0$)	&	Significant	&	Yes	&	Fixed	&	Strongly correlated with the $q$ constant of A($v=0$).	\\
    13	&		&	$\sim$	&	 I($v=1$)	&	Significant	&	Yes	&	Fixed	&	Strongly correlated with the $q$ constant of A($v=0$).	\\
    14	&		&	$\sim$	&	 I($v=2$)	&	Insignificant	&	No	&	-	&		\\
    \colrule	15	&	 A($v=1$)	&	$\sim$	&	 d($v=4$)	&	Insignificant	&	No	&	-	&		\\
    16	&		&	$\sim$	&	 d($v=5$)	&	Insignificant	&	No	&	-	&		\\
    17	&		&	$\sim$	&	 d($v=6$)	&	Insignificant	&	No	&	-	&		\\
    18	&		&	$\sim$	&	 e($v=1$)	&	Insignificant	&	No	&	-	&		\\
    19	&		&	$\sim$	&	 e($v=2$)	&	Insignificant	&	No	&	-	&		\\
    20	&		&	$\sim$	&	 e($v=3$)	&	Insignificant	&	No	&	-	&		\\
    21	&		&	$\sim$	&	 e($v=4$)	&	Insignificant	&	No	&	-	&		\\
    22	&		&	$\sim$	&	 a$'$($v=9$)	&	Insignificant	&	No	&	-	&		\\
    23	&		&	$\sim$	&	 a$'$($v=10$)	&	Insignificant	&	No	&	-	&		\\
    24	&		&	$\sim$	&	 a$'$($v=11$)	&	Insignificant	&	No	&	-	&		\\
    25	&		&	$\sim$	&	 D($v=0$)	&	Insignificant	&	No	&	-	&		\\
    26	&		&	$\sim$	&	 D($v=1$)	&	Insignificant	&	No	&	-	&		\\
    27	&		&	$\sim$	&	 D($v=2$)	&	Insignificant	&	No	&	-	&		\\
    28	&		&	$\sim$	&	 I($v=1$)	&	Insignificant	&	No	&	-	&		\\
    29	&		&	$\sim$	&	 I($v=2$)	&	Insignificant	&	No	&	-	&		\\
    \colrule	30	&	 d($v=3$)	&	$\sim$	&	 e($v=0$)	&	Insignificant	&	No	&	-	&		\\
    31	&		&	$\sim$	&	 e($v=1$)	&	Insignificant	&	No	&	-	&	Negligible weak spin-spin interaction. Strongly correlated with the A($v=0$){}$\sim$ e($v=1$) interaction.\\
    % &		&		&		&		&		&		&	with the A($v=0$){}$\sim$ e($v=1$) interaction.	\\
    32	&		&	$\sim$	&	 e($v=2$)	&	Insignificant	&	No	&	-	&		\\
    33	&		&	$\sim$	&	 a$'$($v=8$)	&	Insignificant	&	No	&	-	&		\\
    34	&		&	$\sim$	&	 a$'$($v=9$)	&	Insignificant	&	No	&	-	&		\\
    \colrule	35	&	 d($v=4$)	&	$\sim$	&	 e($v=0$)	&	Insignificant	&	No	&	-	&	Negligible weak spin-spin interaction. Statistically indeterminable. Floating causes divergency of the fit.\\
    % &		&		&		&		&		&		&	Floating causes divergency of the fit.	\\
    36	&		&	$\sim$	&	 e($v=1$)	&	Insignificant	&	No	&	-	&	Negligible weak spin-spin interaction. Strongly correlated with the A($v=0$){}$\sim$ e($v=1$) interaction.\\
    % &		&		&		&		&		&		&	with the A($v=0$){}$\sim$ e($v=1$) interaction.	\\
    37	&		&	$\sim$	&	 e($v=2$)	&	Insignificant	&	No	&	-	&		\\
    38	&		&	$\sim$	&	 a$'$($v=8$)	&	Insignificant	&	No	&	-	&	Negligible weak spin-spin interaction. Strongly correlated with the A($v=0$){}$\sim$ d($v=4$) interaction and $\lambda$ constant of d($v=4$).\\
    % &		&		&		&		&		&		&	A($v=0$){}$\sim$ d($v=4$) interactio and $\lambda$ constant of d($v=4$).	\\
    39	&		&	$\sim$	&	 a$'$($v=9$)	&	Significant	&	Yes	&	Floated	&		\\
    40	&		&	$\sim$	&	 a$'$($v=10$)	&	Insignificant	&	No	&		&		\\
    \colrule	41	&	 d($v=5$)	&	$\sim$	&	 e($v=1$)	&	Insignificant	&	No	&	-	&	Negligible weak spin-spin interaction. Strongly correlated with the A($v=0$){}$\sim$ e($v=1$) interaction.\\
    % &		&		&		&		&		&		&	with the A($v=0$){}$\sim$ e($v=1$) interaction.	\\
    42	&		&	$\sim$	&	 e($v=2$)	&	Insignificant	&	No	&	-	&		\\
    43	&		&	$\sim$	&	 e($v=3$)	&	Insignificant	&	No	&	-	&		\\
    44	&		&	$\sim$	&	 a$'$($v=9$)	&	Insignificant	&	No	&	-	&		\\
    45	&		&	$\sim$	&	 a$'$($v=10$)	&	Insignificant	&	No	&	-	&		\\
    46	&		&	$\sim$	&	 a$'$($v=11$)	&	Insignificant	&	No	&	-	&		\\
    \colrule	47	&	 a($v=10$)	&	$\sim$	&	 d($v=3$)	&	Insignificant	&	No	&	-	&		\\
    48	&		&	$\sim$	&	 d($v=4$)	&	Significant	&	Yes	&	Fixed	&	Statistically indeterminable.	\\
    49	&		&	$\sim$	&	 e($v=0$)	&	Insignificant	&	No	&	-	&		\\
    50	&		&	$\sim$	&	 e($v=1$)	&	Insignificant	&	No	&	-	&		\\
    51	&		&	$\sim$	&	 a$'$($v=8$)	&	Insignificant	&	No	&	-	&		\\
    52	&		&	$\sim$	&	 a$'$($v=9$)	&	Significant	&	Yes	&	Fixed	&	Statistically indeterminable.	\\
    53	&		&	$\sim$	&	 I($v=0$)	&	Insignificant	&	No	&	-	&		\\
    54	&		&	$\sim$	&	 I($v=1$)	&	Insignificant	&	No	&	-	&		\\
    \colrule	55	&	 a($v=11$)	&	$\sim$	&	 d($v=3$)	&	Insignificant	&	No	&	-	&		\\
    56	&		&	$\sim$	&	 d($v=4$)	&	Significant	&	Yes	&	Floated	&		\\
    57	&		&	$\sim$	&	 d($v=5$)	&	Significant	&	Yes	&	Fixed	&	Strongly correlated with origin and the $A$ constant of a($v=11$).	\\
    58	&		&	$\sim$	&	 e($v=0$)	&	Insignificant	&	No	&	-	&		\\
    59	&		&	$\sim$	&	 e($v=1$)	&	Significant	&	Yes	&	Fixed	&	Floating causes divergency of the fit.	\\
    60	&		&	$\sim$	&	 e($v=2$)	&	Insignificant	&	No	&	-	&		\\
    61	&		&	$\sim$	&	 e($v=3$)	&	Insignificant	&	No	&	-	&		\\
    62	&		&	$\sim$	&	 a$'$($v=8$)	&	Significant	&	Yes	&	Fixed	&	Strongly correlated with origin, $A$, $\lambda$, and $o$ constants of a($v=11$).	\\
    63	&		&	$\sim$	&	 a$'$($v=9$)	&	Significant	&	Yes	&	Floated	&		\\
    64	&		&	$\sim$	&	 a$'$($v=10$)	&	Insignificant	&	No	&	-	&		\\
    65	&		&	$\sim$	&	 I($v=0$)	&	Insignificant	&	No	&	-	&		\\
    66	&		&	$\sim$	&	 I($v=1$)	&	Significant	&	Yes	&	Fixed	&	Strongly correlated with the $o$ constant of a($v=11$).	\\
    67	&		&	$\sim$	&	 I($v=2$)	&	Insignificant	&	No	&	-	&		\\
    68	&		&	$\sim$	&	 D($v=0$)	&	Significant	&	Yes	&	Fixed	&	Statistically indeterminable.	\\
    69	&		&	$\sim$	&	 D($v=1$)	&	Insignificant	&	No	&	-	&		\\
    \colrule	70	&	 a($v=12$)	&	$\sim$	&	 d($v=4$)	&	Insignificant	&	No	&	-	&		\\
    71	&		&	$\sim$	&	 d($v=5$)	&	Insignificant	&	No	&	-	&		\\
    72	&		&	$\sim$	&	 e($v=2$)	&	Insignificant	&	No	&	-	&		\\
    73	&		&	$\sim$	&	 e($v=3$)	&	Insignificant	&	No	&	-	&		\\
    74	&		&	$\sim$	&	 a$'$($v=9$)	&	Insignificant	&	No	&	-	&		\\
    75	&		&	$\sim$	&	 a$'$($v=10$)	&	Insignificant	&	No	&	-	&		\\
    76	&		&	$\sim$	&	 D($v=1$)	&	Insignificant	&	No	&	-	&		\\
    77	&		&	$\sim$	&	 D($v=2$)	&	Insignificant	&	No	&	-	&		\\
    78	&		&	$\sim$	&	 I($v=1$)	&	Insignificant	&	No	&	-	&		\\
    79	&		&	$\sim$	&	 I($v=2$)	&	Insignificant	&	No	&	-	&		\\
    \colrule	80	&	 a$'$($v=8$)	&	$\sim$	&	 e($v=0$)	&	Insignificant	&	No	&	-	&		\\
    81	&		&	$\sim$	&	 e($v=1$)	&	Insignificant	&	No	&	-	&	Negligible weak spin-spin interaction. Statistically indeterminable.	\\
    \colrule	82	&	 a$'$($v=9$)	&	$\sim$	&	 e($v=0$)	&	Insignificant	&	No	&	-	&		\\
    83	&		&	$\sim$	&	 e($v=1$)	&	Insignificant	&	No	&	-	&		\\
    84	&		&	$\sim$	&	 e($v=2$)	&	Insignificant	&	No	&	-	&		\\
    \colrule	85	&	 a$'$($v=10$)	&	$\sim$	&	 e($v=1$)	&	Insignificant	&	No	&	-	&		\\
    86	&		&	$\sim$	&	 e($v=2$)	&	Insignificant	&	No	&	-	&		\\
    87	&		&	$\sim$	&	 e($v=3$)	&	Insignificant	&	No	&	-	&		\\
    \colrule															
    \colrule															

  \end{longtable}
\end{ThreePartTable}

\begin{table}
\caption{Term values of $B(0)$\textsuperscript{$a$}.}
  \label{B-term}
\footnotesize
\begin{threeparttable}
\begin{tabular}{cl}
\colrule
$J$                & \multicolumn{1}{c}{$F_{1e}$}           \\
\colrule
  0                & 86\,917.355(2)     \\
  1                & 86\,920.896(1)     \\
  2                & 86\,927.976(1)     \\
  3                & 86\,938.5946(9)    \\
  4                & 86\,952.7505(8)    \\
  5                & 86\,970.4474(8)    \\
  6                & 86\,991.6796(7)    \\
  7                & 87\,016.4484(7)    \\
  8                & 87\,044.7532(7)    \\
  9                & 87\,076.5925(7)    \\
 10                & 87\,111.9658(7)    \\
 11                & 87\,150.8695(8)    \\
 12                & 87\,193.3081(8)    \\
 13                & 87\,239.2712(8)    \\
 14                & 87\,288.7648(8)    \\
 15                & 87\,341.7827(9)    \\
 16                & 87\,398.3224(9)    \\
 17                & 87\,458.344(1)     \\
 18                & 87\,521.963(1)     \\
 19                & 87\,589.071(1)     \\
 20                & 87\,659.682(1)     \\
 21                & 87\,733.807(1)     \\
 22                & 87\,811.440(1)     \\
 23                & 87\,892.579(1)     \\
 24                & 87\,977.224(2)     \\
 25                & 88\,065.364(2)     \\
 26                & 88\,157.007(2)     \\
 27                & 88\,252.143(2)     \\
 28                & 88\,350.760(2)     \\
 29                & 88\,452.863(2)     \\
 30                & 88\,558.455(2)     \\
 31                & 88\,667.510(2)     \\
 32                & 88\,780.049(2)     \\
 33                & 88\,896.067(3)     \\
 34                & 89\,015.545(3)     \\
 35                & 89\,138.495(3)     \\
 36                & 89\,264.866(4)     \\
 37                & 89\,394.705(4)     \\
 38                & 89\,527.998(5)     \\
 39                & 89\,664.728(5)     \\
 40                & 89\,804.886(6)     \\
 41                & 89\,948.482(7)     \\
 42                & 90\,095.482(9)     \\
 43                & 90\,245.93(1)      \\
 44                & 90\,399.79(1)      \\
 45                & 90\,557.05(1)      \\
 46                & 90\,717.74(2)      \\
 47                & 90\,881.76(2)      \\
 48                & 91\,049.19(3)      \\
 49                & 91\,220.05(3)      \\
 50                & 91\,394.20(4)      \\
\colrule
\end{tabular}
\begin{tablenotes}
\begin{footnotesize}
\item[$a$] In units of \wn{} and with $1\sigma$ fitting uncertainties given in parentheses in units of the least-significant digit that are additional to a 0.01\,\wn{} systematic uncertainty.
\end{footnotesize}
\end{tablenotes}
\end{threeparttable}
\end{table}

\clearpage

Further the Pgopher file including the entire data set and fitting routine of the deperturbation analysis as well as the data of Figures \ref{fig:BA00} and \ref{fig:VUV A-X spectrum} in digital format are provided.

\end{document}